\documentclass[fleqn,12pt]{article}
\usepackage{amsmath,amsfonts,amssymb,latexsym}
\usepackage{graphicx}
\usepackage[nospace]{cite}
\usepackage{psfrag}
\usepackage{lscape}

\usepackage[sc,small]{caption2}
\setcaptionwidth{14cm}

\makeatletter
\newbox\slashbox \setbox\slashbox=\hbox{$/$}
\def\pFMslash#1{\setbox\@tempboxa=\hbox{$#1$}
  \@tempdima=0.5\wd\slashbox \advance\@tempdima 0.5\wd\@tempboxa
  \copy\slashbox \kern-\@tempdima \box\@tempboxa}

\makeatother

\newcommand{\Ge}{\epsilon}
\newcommand{\Geps}{\varepsilon}

\newcommand{\GTh}{\Theta}

\newcommand{\cM}{{\scriptscriptstyle\cal M}}
\newcommand{\cN}{{\scriptscriptstyle\cal N}}
\newcommand{\cK}{{\scriptscriptstyle\cal K}}
\newcommand{\cL}{{\scriptscriptstyle\cal L}}
\newcommand{\cP}{{\scriptscriptstyle\cal P}}

\newcommand{\CM}{{\cal M}}

\newcommand{\CL}{{\cal L}}
\newcommand{\CO}{{\cal O}}

\newcommand{\CS}{{\cal S}}

\newcommand{\CV}{{\cal V}}
\newcommand{\CZ}{{\cal Z}}

\newcommand{\ft}[2]{{\textstyle {\frac{#1}{#2}} }}
\newcommand{\dd}{\partial}

\newcommand{\be}{\begin{equation}}
\newcommand{\ee}{\end{equation}}
\newcommand{\ben}{\begin{displaymath}}
\newcommand{\een}{\end{displaymath}}
\newcommand{\bea}{\begin{eqnarray}}
\newcommand{\eea}{\end{eqnarray}}
\newcommand{\ba}{\begin{eqnarray}}
\newcommand{\ea}{\end{eqnarray}}

\newcommand{\bean}{\begin{eqnarray*}}
\newcommand{\eean}{\end{eqnarray*}}

\def\moth{\mathsurround=0pt}
\newdimen\zo \zo=0pt

\def\tick{\leaders\hrule height 0.5ex depth 0pt \hskip 0.5pt}
\def\upboxfill{$\moth \setbox\zo\hbox{\tick}%
  \hskip 2pt\hbox to 0pt{$\tick$\hss}\hrulefill \hbox to
6pt{$\tick$\hss}$}

\def\dtick{\leaders\hrule height .34pt depth .5ex \hskip 0.5pt}
\def\downboxfill{$\moth \setbox\zo\hbox{\dtick}%
  \hskip 2pt\hbox to 0pt{$\dtick$\hss}\hrulefill \hbox to
6pt{$\dtick$\hss}$}

\newcommand{\la}{\label}

\newcommand{\Ref}[1]{(\ref{#1})}

\makeatletter
\@addtoreset{equation}{section}
\makeatother

\newcommand{\SO}[1]{{{\rm SO}({#1})}}

\begin{document}

\thispagestyle{empty}

\begin{center}
AEI-2006-098 \hspace{1cm}
ENSL-00122693 \hspace{1cm}
ITP-UU-06/55 \hspace{1cm}
SPIN-06/45
 \end{center}

\vspace*{.4cm}


\begin{center}
{\bf\LARGE Holography of \\ [4mm]
D-Brane Reconnection}
\bigskip\bigskip

{\bf Marcus~Berg\footnotemark,
\footnotetext{Marcus.Berg@aei.mpg.de}
Olaf Hohm\footnotemark,
\footnotetext{O.Hohm@phys.uu.nl}
and
Henning~Samtleben\footnotemark}
\footnotetext{Henning.Samtleben@ens-lyon.fr}

\vspace{.3cm}
$^1${\em }
{\em Max Planck Institute for Gravitational Physics}  \\
{\em Albert Einstein Institute}\\ M\"uhlenberg 1, D-14476 Potsdam, Germany
\
\vspace{.5cm}

$^2${\em}
{\em Spinoza Institute and Institute for Theoretical Physics} \\ Leuvenlaan 4, 3584 CE Utrecht, The Netherlands
\
\vspace{.5cm}

$^3${\em}
{\em Laboratoire de Physique, ENS-Lyon} \\
46, all\'ee d'Italie, F-69364 Lyon CEDEX 07, France

\end{center}
\renewcommand{\thefootnote}{\arabic{footnote}}
\setcounter{footnote}{0}
\bigskip
\medskip
\begin{abstract}
Gukov, Martinec, Moore and Strominger found that the D1-D5-D5$'$
system with the D5-D5$'$ angle at 45 degrees admits a
deformation $\rho$ preserving supersymmetry. Under this
deformation, the D5-branes and D5'-branes reconnect along a single
special Lagrangian manifold.
We construct the near-horizon limit of this brane setup (for which no
supergravity solution is currently known), imposing the requisite
symmetries perturbatively in the deformation $\rho$.
Reducing to the three-dimensional effective gauged supergravity,
we compute the scalar potential and verify the
presence of a deformation with the expected properties.
We compute the conformal dimensions as
functions of $\rho$. This
spectrum
naturally organizes into ${\cal N}=3$ supermultiplets, corresponding
to the 3/16 preserved by the brane system.
We give some remarks on the symmetric orbifold CFT for
$Q_{\rm D5}=Q_{{\rm D5}'}$, outline the computation of $\rho$-deformed correlators in
this theory, and probe computations in our $\rho$-deformed background.

\end{abstract}

\renewcommand{\thefootnote}{\arabic{footnote}}
\vfill \leftline{{December 2006}}

\setcounter{footnote}{0}

\newpage
\setcounter{page}{1}
\renewcommand{\thepage}{\roman{page}}

{\baselineskip 13pt \tableofcontents}

\newpage
\renewcommand{\thepage}{\arabic{page}}
\setcounter{page}{1}

\section{Introduction}

The D1-D5 system has been a target of much interest ever since the classic
Strominger-Vafa paper on its relation to black hole entropy
\cite{Strominger:1996sh}, and its later inception in the AdS/CFT
correspondence. The worldsheet theory on the D1-brane --- the boundary
theory
in AdS/CFT --- is an ${\cal N}=(4,4)$ superconformal theory.
It is textbook knowledge
\cite[Ch.\ 11.1]{Polchinski:1998rr} that in addition to the ${\cal N}=4$ superconformal algebra,
that has an $SU(2)$ bosonic subalgebra, there is the
``large"  ${\cal N}=4$ algebra, with an $SU(2) \times SU(2)$ subalgebra. The
place of this
enlarged symmetry in the AdS/CFT correspondence, and how to break it,
is the subject of this paper. Breaking this
enlarged symmetry may ultimately help understanding 
deformations of
the original D1-D5 system broken down to ${\cal N}=(3,3)$ supersymmetry.

\begin{figure}[htb]
\label{fig:branes}
\begin{center}
\psfrag{a}[bc][bc][.8][0]{$a)$}
\psfrag{b}[bc][bc][.8][0]{$b)$}
\psfrag{c}[bc][bc][.8][0]{$c)$}
\psfrag{d}[bc][bc][.8][0]{$d)$}
\psfrag{45}[bc][bc][.8][0]{$45^{\circ}$}
\psfrag{M}[bc][bc][.7][0]{${\cal M}$}
\includegraphics[width=0.7\textwidth]{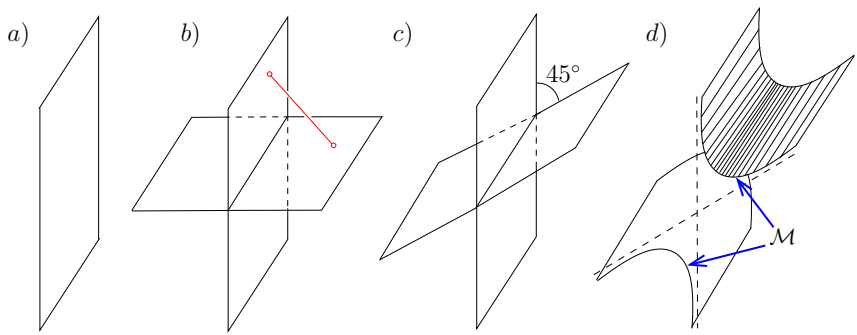}
\caption{$a)$ D1-D5 system, for comparison. The D1-brane is delocalized
on the D5-brane. $b)$ Orthogonal D1-D5-D5$'$ system with string inducing
nonlocal interactions on D1-brane.
$c)$ Tilted  D1-D5-D5$'$ system. $d)$ $\rho$-deformed D1-D5-D5$'$
system on the curved manifold ${\cal M}$ times the 1+1 intersection.
As will be explained later, ${\cal M}$ is simply connected, despite
appearances.}
\vspace{-5mm}
\end{center}
\end{figure}

Already in 1999, de Boer, Pasquinucci and Skenderis
\cite{deBoer:1999rh}
(based on earlier work~\cite{Boonstra:1998yu,Elitzur:1998mm}) 
initiated the study of AdS/CFT dual pairs with large ${\cal N}=(4,4)$ symmetry, and found the
requisite solution of Type IIB supergravity: two orthogonal D5-brane stacks, intersecting
over a D-string. This is the D1-D5-D5$'$ system (fig.\ \ref{fig:branes}b).
Compared to the D1-D5 system,
the $SO(4)$ symmetry of transverse rotations of the single D5-brane stack is
doubled to $SO(4) \times SO(4)$ to include independent transverse rotations of both D5-brane
stacks. In the
near-horizon limit, the boundary theory on the D1-branes should be the
large ${\cal N}=(4,4)$ theory. The D1-D5-D5$'$ system received some further attention
\cite{Berg:2001ty,Berg:2002hy}, but it remains mysterious;  
simply decreasing the ratio of D5$'$ to D5 charge, one does not arrive at the D1-D5
system, and the worldsheet theory is nonlocal~\cite{deBoer:1999rh},
due to D5-D5$'$ interactions (the open string shown in fig. \ref{fig:branes}b).

Tilting the branes gave a new perspective on this.
In the first of a series of three papers
 \cite{Gukov:2004ym,Gukov:2004id,Gukov:2004fh},
 Gukov, Moore, Martinec and Strominger (GMMS) 
found\footnote{Earlier related work on intersecting branes in M-theory 
includes \cite{Gauntlett:1997pk,Gauntlett:1998vk,Gauntlett:1998kc}.}
that if the D5$'$-branes are tilted at
$45^{\circ}$ relative to the D5-branes as shown in fig.\
\ref{fig:branes}c (breaking supersymmetry to
3/16), a new possibility appears; there
is a supersymmetric noncompact manifold ${\cal M}$ that the branes
can reconnect along, as in fig.\ \ref{fig:branes}d.
GMMS identified a deformation called
$\rho$ that would describe this reconnection, and pointed out that it
would break the rotational symmetry to the diagonal.

Although the deformed system is less symmetric, the
complications of the D1-D5-D5$'$ system that come 
from nonlocal interactions should not arise when D5-branes and
D5$'$-branes are joined on ${\cal M}$,
and in fact the $\rho$-deformed system may have more in common with the single D1-D5
system (e.g.\ its Higgs branch \cite{Gukov:2004ym}) than
the undeformed system did.

The corresponding deformed 10-dimensional solution with the branes extended
along the curved manifold ${\cal M}$ is not known explicitly. In section~3 we
construct its near-horizon limit (eq.~\Ref{defbein}) by imposing the requisite
symmetries perturbatively in the deformation parameter $\rho$.
Identifying the field $\rho$ in the Kaluza-Klein
spectrum of fluctuations around the undeformed $AdS_{3}\times S^{3}\times S^{3}\times S^{1}$
background, we reduce to the three-dimensional effective supergravity
with $SO(4) \times SO(4) \rightarrow SO(4)_{\rm diag}$ gauge symmetry
breaking.
In section~5, we compute the scalar potential and show the
(from a supergravity point of view somewhat surprising)
presence of a flat ``valley'', i.e.\ a deformation marginal to all
orders, shown in Figure~\ref{fig:valley}.
We verify that evolution of the scalars along the valley
reproduces all the properties expected for the $\rho$-deformation
and in particular breaks supersymmetry to ${\cal N}=(3,3)$.
We compute the conformal dimensions $\Delta$ in the theory as
functions of $\rho$, by computing the spectrum along the flat valley. This
spectrum naturally organizes into ${\cal N}=3$ supermultiplets.
In section~6, we give some remarks on the symmetric orbifold CFT that was
conjectured in \cite{deBoer:1999rh,Gukov:2004ym} 
to describe at least some aspects of the deformed 
boundary theory for $Q_{\rm D5}=Q_{{\rm D5}'}$. We outline the computation
 of $\rho$-deformed correlators in
this theory, and probe computations in the $\rho$-deformed background.

To some readers,
D-brane reconnection will be more  familiar
\cite[Ch.\ 13.4]{Polchinski:1998rr} in the context of D-branes at angles,
where below a certain critical angle, a tachyonic mode develops and
the branes move apart.
Clearly this is quite different from the marginal deformation we are
interested in,
where the reconnected branes can be disconnected again
by sending $\rho \rightarrow 0$, at no cost in energy (see also \cite{Hashimoto:2003pu}).
A more closely related kind
of brane reconnection along special Lagrangian manifolds has been extensively
studied in the literature in other contexts,
like \cite{Gukov:2002es,Lambert:2002ms} for
M-theory on $G_2$ manifolds.

\section{The D1-D5-D5$'$ system intersecting at angles}
We begin with a review of
the D1-D5-D5$'$ system
with $SO(4) \times SO(4)'$ symmetry
and explain how its near-horizon limit
$AdS_3 \times S^3 \times S^3 \times S^1$ arises. The boundary theory of
the $AdS_3$ factor has large ${\cal N}=(4,4)$ supersymmetry in $1+1$ dimensions.
The supergravity solution for this system
was studied for orthogonal intersection in
 \cite{deBoer:1999rh},
for D5-D5$'$ intersecting at angles
in \cite{Gauntlett:1997pk}, and
for D1-D5-D5$'$ intersecting at angles
in \cite{Gukov:2004ym}.
Here we summarize the results without derivation.

We denote the 10-dimensional coordinates by
$\{t,z,x_1,x_2,x_3,x_4,y_1,y_2,y_3,y_4\}$, and number them
by $0,1,\ldots,9$. We will often use the notation
$x^2=x_1^2+\ldots+x_4^2$, $y^2=y_1^2+\ldots+y_4^2$. The angles
between D5 and D5$'$-branes are denoted $\theta_{26}$ for the
angle in the $x_1-y_1$ plane, and so on. For D5-branes
intersecting at equal angles in all four planes
$\theta_{26}=\theta_{37}=\theta_{48}=\theta_{59}=:\theta$,
 we have the supersymmetry conditions
 \cite{Gauntlett:1997pk}
\be
\Gamma_{012345} \epsilon_{\rm R} = \epsilon_{\rm L} \; , \quad
\exp(\theta(\Gamma_{26}+\Gamma_{37}+\Gamma_{48}+\Gamma_{59}))
\epsilon_{\rm L,R} = \epsilon_{\rm L,R} \;,
\ee
that together preserve 6 Killing spinors, i.e.\ 3/16 supersymmetry.
Including D1-branes delocalized in the $\vec{x}$ and $\vec{y}$
directions adds the condition
\be
\Gamma_{01} \epsilon_{\rm R} = \epsilon_{\rm L}  \; ,
\ee
that further breaks supersymmetry from 3/16 to 1/16,
again at nonzero $\theta$.
The Type IIB supergravity solution for angle $\theta$ is
\ba
ds^2 &=& (H_1^{(+)}H_1^{(-)} \det U)^{-1/2} (-dt^2+dz^2)
+\sqrt{H_1^{(+)}H_1^{(-)}}{U_{11} \over \sqrt{\det U}}
(d\vec{x})^2 \\ \nonumber
&&+ \sqrt{H_1^{(+)}H_1^{(-)}}{U_{22} \over \sqrt{\det U}}
(d\vec{y})^2
+{2 U_{12} \over \sqrt{\det U}} d\vec{x} \cdot d\vec{y}
\;,
\ea
with RR 3-form field strength
\be
F_3 = dt \wedge dz \wedge d(H_1^{(+)}H_1^{(-)})^{-1}
+*_x dU_{11}+*_y dU_{22}
\;,
\label{3form}
\ee
and dilaton
\be
e^{-2\phi}= {1 \over g_{\rm s}^2}{\det U \over H_1^{(+)}H_1^{(-)}}
\;,
\ee
where the harmonic functions $H_1^{(+)}$, $H_1^{(-)}$ are 
\be
H_1^{(+)} = 1 + {g_{\rm s} Q_1 \over x^2}
\; , \qquad
H_1^{(-)} =  1 + {g_{\rm s} Q_1 \over y^2}
\;,
\ee
and the matrix $U$ is given by
\be
U =
\left( \begin{array}{cc}
\csc \theta+{\textstyle g_{\rm s} Q_5^+ \over {\textstyle x^2}}
&  -\cot \theta \\ [2mm]
-\cot \theta & \csc \theta+ {\textstyle g_{\rm s} Q_5^- \over
  {\textstyle y^2
}}
\end{array} \right) \; .
\ee
Here $Q_1$ is the D1-brane charge, $Q_5^+$  is the D5-brane charge,
and $Q_5^-$ is the D5$'$-brane charge.
We set $Q_5^+=Q_5^-=:Q $ for reasons discussed in \cite{deBoer:1999rh,Gukov:2004ym}.
The D1-brane charge $Q_1$ is fixed in terms
of the D5-brane charges as $Q_1 = {L \over 4\pi^2}Q_5$.

The near-horizon limit of this solution was studied
in the above papers
\cite{deBoer:1999rh,Gukov:2004ym,Gauntlett:1998kc}.
Somewhat counterintuitively, the near-horizon
limit is the same regardless of the rotation angle~$\theta$.
It is, setting $g_s=1$ and $L={4\pi^2}\tilde{L}$,
\bea
ds^2 &=& {x^2 y^2 \over \tilde{L} Q^2} (-dt^2 + dz^2)
+Q \tilde{L}\left({dx^2 \over x^2} + d\Omega_+^2\right)
+ Q\tilde{L}\left({dy^2 \over y^2} + d\Omega_-^2\right) \; .
\eea
After a change of coordinates
\ba
r &=& (\sqrt{2} Q^{-1/2} )  xy\;,\qquad
\phi ~=~ {1 \over \sqrt{2}} \log{y \over x} \; ,    \label{NHtrans}
\ea
using that $dx^2/x^2+dy^2/y^2= a\,  dr^2/r^2+b \, d\phi^2$
for some constants $a$ and $b$, and identifying\footnote{ This identification was performed in 
\cite{deBoer:1999rh} and criticized in \cite{Gukov:2004ym}.
We will have nothing new to say about this.}
 $\phi$ to make an $S^1$,
the near-horizon metric
becomes
\be
ds^2 = {r^2 \over 2R^2} (-dt^2 + dz^2) +{R^2 \over 2r^2} dr^2
+ R^2 d\Omega_+^2 + R^2 d\Omega_-^2 +  R^2 d\phi^2
\;,
\label{AdSSSS}
\ee
with $R^2=Q\tilde{L}$,
which we recognize as $AdS_3 \times S^3 \times S^3 \times S^1$.
The matter fields are
\ba
F_3 &=& {\rm vol}(AdS_3) + {\rm vol}(S_+^3)+ {\rm vol}(S_-^3) \;,\qquad
e^{\phi} ~=~ 1/ \tilde{L}^2
\;.
\ea

Here, we have ${\cal N}=(4,4)$ supersymmetry. With the solution at
hand, it is fairly clear from (\ref{NHtrans}) that the new
coordinate system $(r,\phi)$ moves along with $\theta$. Also,
$\theta$ only occurs in the constant part of the matrix $U$, 
which is neglected in the near-horizon limit. We see that the 
dependence on the angle $\theta$ will be lost in the near-horizon
limit.

Going back to the full solution, it is interesting to consider whether
instead of flat worldvolumes, there are any natural supersymmetric manifolds
the branes could extend along. One 
 suitable manifold was found in~\cite{Gukov:2004ym}, as we now discuss.

\subsection{Brane reconnection}
\label{sec:brane}

\begin{figure}[h]
\begin{center}
\psfrag{x}[bc][bc][1][0]{$x$}
\psfrag{y}[bc][bc][1][0]{$y$}
\includegraphics[width=0.3\textwidth,height=4.5cm]{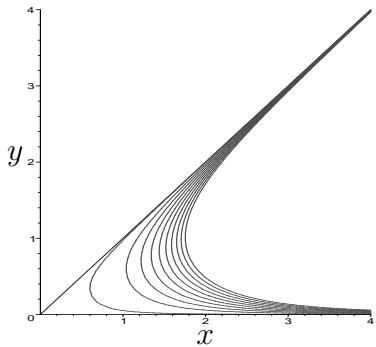}
\end{center}
\vspace{-5mm}
\caption{$\rho=xy(x^2-y^2)$ for various values of $\rho$.}
 \label{figrho1}
\end{figure}

\begin{figure}[ht]
\begin{center}
\includegraphics[width=0.6\textwidth,height=7cm]{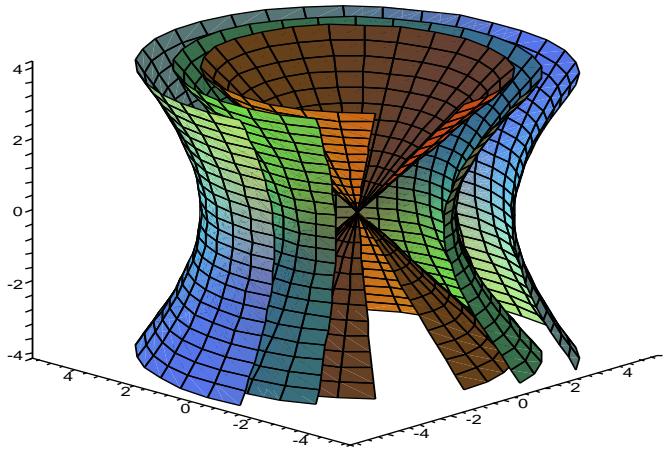}
\end{center}
\vspace{-5mm}
\caption{Cylindrical plot $[r,\theta,\sigma]$
from eq.~\Ref{rhosigmaxy} for various values of $\rho$.
The surface at the center represents $\rho=0$. The apparent disconnectedness
in fig.~\ref{fig:branes} can now be understood as
a hyperbolic ``conic section" of the connected $\rho>0$ manifold. }
\label{rhoplot2}
\end{figure}

Let ${\bf C}^m$ have complex coordinates
$z_k = x_k + i y_k$, $k=1,...,m$. Consider the submanifold
 given by
\be
{\rm Im} \, z^n = {\rm const} \;,
\label{Meqn}
\ee
where $z=|x_k|+i |y_k|$.
It is special Lagrangian for $m=n$
\cite{Harvey:1982xk}.
For the system described in the previous section, the nonintersecting part of the D5-branes
is  ${\bf R}^4 \times {\bf R}^4
\simeq {\bf C}^4$, so combining the two lengths
$x^2=x_1^2+\ldots+x_4^2$ and
$y^2=y_1^2+\ldots+y_4^2$
into a complex number\footnote{Note that this complex $z$
has nothing to do with the coordinate $z$ in the previous section.} $z=x+iy$,
we could consider letting the D5-branes extend along on the manifold \Ref{Meqn}
for $n=4$. In particular, let
\be
\rho := {1 \over 4}\, {\rm Im}\,  z^4 = xy (x^2-y^2) \;,
\label{rhoeq}
\ee
then we can define the noncompact special Lagrangian manifold $\CM$ as \cite{Gukov:2004ym}
\be
\CM = \Big\{ \rho=\mbox{constant} \; , \; {x_i \over x} = {y_i \over y} \Big\}
\quad i = 1,..,4 \; .
\ee
We plot \Ref{rhoeq} in Fig.\ \ref{figrho1} for a few values of $\rho$.
Note that the first quadrant is all there is, since $x$ and $y$
are lengths.
Defining $\sigma=(1/4)\, {\rm Re }\, z^4$, and noting that
\be
\rho^2+\sigma^2 = {(x^2 + y^2)^4 \over 16} =: r^2
\;,
\label{rhosigmaxy}
\ee
we can go to cylindrical coordinates $[r,\theta,\sigma]$ 
with $\sigma$ as the vertical axis.
A plot in these coordinates appears in Fig.\ \ref{rhoplot2}.
For $\rho>0$, the full manifold has topology
${\bf R} \times {\bf S}^3$,
and this is represented by ${\bf R} \times {\bf S}^1$ in Figure \ref{rhoplot2}.
We see clearly that for $\rho=0$, the manifold
degenerates to two separate branches
${\bf R}^4 \times {\bf R}^4$, represented by ${\bf R}^2 \times {\bf
  R}^2$ in the figure (the conical singularity is an artifact of the embedding).
We do not know of a supergravity solution for D5-branes
extended along the noncompact special Lagrangian manifold $\CM$.

Importantly, if we take the near-horizon limit
of the solutions with the branes wrapped on \Ref{rhoeq},
we expect the symmetry to be reduced.
In the undeformed brane system,
the $SO(4) \times SO(4)'$ symmetry corresponds to
independent transverse rotations of the two sets of D5-branes.
After deformation, we see in Fig.\ \ref{rhoplot2} that transverse
rotations are no longer independent; the two sets of branes are now
wrapping a single connected manifold, with the symmetry
broken as \cite{Gukov:2004ym}
\be
SO(4) \times SO(4)' \quad \stackrel{\rho>0}{\longrightarrow} \quad SO(4)_{\rm diag}
\;.
\label{symred}
\ee
This is intriguing; through this deformation, one could hope to make
progress in connecting the D1-D5-D5$'$ system to
the system of (presently) greater physical interest, the single D1-D5
system. Since the deformation is only an ${\cal N}=3$ modulus and not an ${\cal N}=4$
modulus --- the deformation only exists in the tilted system, not the
orthogonal system --- this will primarily apply to a version of the
D1-D5 system broken to ${\cal N}=3$ supersymmetry.
We will comment on this in the Conclusions.

A valuable clue for understanding the $\rho$ deformation by holography
is parity. In \Ref{rhoeq}, $\rho$ is odd under interchange of the two three-spheres
($x \leftrightarrow y$). This will be important in each of the
following sections.\footnote{The paper \cite{Yamaguchi:1999gb} even
  studied an orbifold under this parity. 
That orbifold preserves different symmetries
than we are interested in here.}

Even though we do not have a full deformed supergravity solution,
we can attempt to model its near-horizon limit by
imposing the symmetry reduction~\Ref{symred}.
If the boundary theory is to remain a CFT,
we can try to look for deformations that leave the $AdS_{3}$ part
of the dual geometry untouched.
As we will see in the next section, such
a deformation exists and it breaks supersymmetry to ${\cal N}=(3,3)$.

\section{The deformed near-horizon limit}
\label{sec:NH}

In this section we construct the near-horizon limit
of the (currently unknown) $\rho$-deformed brane solution perturbatively in
the deformation parameter $\rho$ by exploiting the symmetries
preserved by the deformation.

Let us first consider what simplifications we can  impose on the
Type IIB  field equations.
If we  want a deformed solution where the dilaton is still constant,
the source 
$F^{MNK}F_{MNK}$
in the dilaton equation of motion will have to stay zero.
In the undeformed case, this was ensured by tuning the three-form
fluxes~\Ref{3form} as
\bea
F_{mnk}F^{mnk}+F_{\bar{m}\bar{n}\bar{k}}F^{\bar{m}\bar{n}\bar{k}}
&=&  \frac8{g^{2}}\;,
\label{Hconst}
\eea
so this must still hold 
for the deformed flux.
The remaining nontrivial field equations are
 \bea\label{IIBeq}
  R_{AB}=\frac{1}{4}F_A^{\hspace{0.5em}CD}F_{BCD}\;, \qquad
  D_A F^{ABC}=0\; .
 \eea
Then we make the  ten-dimensional metric ansatz
\be
ds^2 = ds_{AdS{}_3}^2
+ ds_6^2 +  R^2 d\phi^2 \; ,
\ee
where $ds_6^2$ denotes the deformation of $S^3 \times S^3$
that we will now construct.
(The answer is given in ~\Ref{defbein} below).
Note that as previously stated, these equations are nontrivial only along the
sphere coordinates but leave the $AdS_3$ part intact.

We now summarize the symmetries we want to impose.
Recall that the isometry group of the undeformed background
$AdS_{3}\times S^{3}\times S^{3}$ is given by
\bea
G_{\rm iso}&=&
SO(2,2)\times SO(3)_{L}\times SO(3)_{L}' \times SO(3)_{R}\times SO(3)_{R}'
\;.
\label{iso}
\eea
Here $SO(2,2)=SL(2)_{L}\times SL(2)_{R}$ describes the $AdS_{3}$ isometries
and $SO(4)\equiv SO(3)_{L} \times SO(3)_{R}$ and
$SO(4)'\equiv SO(3)_{L}' \times SO(3)_{R}'$
constitute the isometry groups of the two spheres.
The subscripts L,R, on the other hand, correspond to
the splitting into left- and right-movers in the two-dimensional
boundary CFT.
Out of the $SO(3)$ factors in~(\ref{iso}),
the deformation preserves only the diagonal subgroup
\bea
SO(4)^{(D)}&=&
SO(3)^{(D)}_L\times SO(3)^{(D)}_R
\nonumber\\[1ex]
&\equiv&
{\rm diag}\Big(SO(3)_{L}\times SO(3)_{L}'\Big)
\times
{\rm diag}\Big(SO(3)_{R}\times SO(3)_{R}'\Big)
\;.
\label{isodef}
\eea
The solution can thus be constructed
in terms of those $S^{3}$ sphere harmonics that are left invariant by the
corresponding diagonal combinations of  Killing vector fields.
Moreover, as pointed out in the previous section,
the deformation parameter $\rho$ is odd under exchange of the two spheres.
Together with the invariance requirements this puts  very strong restrictions
on the deformed solution. 

To make this manifest, we need to introduce a little more notation. We
parametrize the upper hemispheres of the spheres by coordinates
$x^m$ and $y^{\bar{m}}$, which are simply the projections of the
Cartesian coordinates of the embedding space~${\bf R}^4$,
 \bea
  X^{\hat{A}}=(x^m,\sqrt{1-x^2})\;, \qquad
  Y^{\hat{A}}=(y^{\bar{m}},\sqrt{1-y^2})\;,
  \label{XAYA}
 \eea
with $x^{2}=\sum_i (x^i)^2$, $y^{2}=\sum_i (y^i)^2$.
The sphere metrics in these coordinates
are given by
 \bea\label{sphere}
  g_{mn}=\delta_{mn} + \frac{x_m x_n}{1-x^2}\;,\qquad
   g_{\bar{m}\bar{n}}=\delta_{\bar{m}\bar{n}} + \frac{y_{\bar{m}} y_{\bar{n}}}{1-y^2}\;.
 \eea
The $SO(4)$ isometries on the first sphere are generated by 6 Killing
vectors $K_{L}\,{}_{(k)}^{\hspace{0.3em}i}$, $K_{R}\,{}_{(k)}^{\hspace{0.3em}i}$, which read
 \bea
  K_{L,R}\,{}_{(k)}^{\hspace{0.3em}i} = -\frac{1}{2}\left(\varepsilon^i_{\hspace{0.3em}
  km}x^m \pm \delta^i_k\sqrt{1-x^2}\right)\;,
 \eea
and similarly for $SO(4)'$.
The normalisation is chosen such that
the Lie brackets close according to the standard $SO(3)$ algebra:
 \bea\label{su2structure}
  [K_{L\,(a)},K_{L\,(b)}]&=&\varepsilon_{ab}^{\hspace{0.8em}c}K_{L\,(c)}\;,\qquad
  \mbox{etc.}
 \eea
The computations are significantly simplified by use of the
vielbein formalism. A convenient $SO(3)$ frame is given
by either half of the Killing vectors themselves, e.g.\ the $K_L$:
 \bea\label{so3frame}
  e_a^{\hspace{0.2em}m}(x):=2 K_L\,{}_{(a)}^{\hspace{0.3em}m}(x)\;, \qquad
  e_{\bar{a}}^{\hspace{0.2em}\bar{m}}(y):=2 K_{L}\,{}_{(\bar{a})}^{\hspace{0.3em}\bar{m}}(y)\;.
 \eea

Because of the algebra~\Ref{su2structure}, invariance under the diagonal
combinations~\Ref{isodef}
of Killing vector fields reduces to invariance under the three combinations
 \bea\label{diagiso}
  K_{D\,(k)}\equiv\sqrt{1-x^2}\frac{\partial}{\partial x^k} + \sqrt{1-y^2}
  \frac{\partial}{\partial y^{\bar{k}}}\;,
 \eea
which upon commutation generate the full $SO(4)^{(D)}$.
We will now construct these invariant sphere harmonics.

Let us start with a ten-dimensional scalar field and
consider its full Kaluza-Klein expansion~(\ref{ExpansionScalar})
in terms of $S^{3}\times S^{3}$ sphere functions $X^{[j,j]}(x)$, $Y^{[j',j']}(y)$
labeled by their spins $j$, $j'$. Under the diagonal $SO(4)_{D}$
this expansion contains an infinite number of singlet excitations,
namely one in each product $X^{[j,j]}Y^{[j',j']}$ for $j=j'$,
corresponding to the decomposition
$[j,j;j,j]\rightarrow [0,0]+\dots$
under the diagonal $SO(4)_{D}$.
Explicitly, this corresponds to a truncation of \Ref{ExpansionScalar}
to an expansion
\bea
\Phi(z,x,y) &=& \sum_{j} \;\varphi_{j}(z)\,u^{2j}\;,
\label{Uexpansion}
\eea
 where $u$ is the inner
product of (\ref{XAYA}) in the embedding space:
 \bea
  u~\equiv~ X^{\hat{A}}Y^{\hat{A}} = \sum_m x^m y^{\bar{m}}
  + \sqrt{1-x^2}\sqrt{1-y^2}
 \;.
 \eea
One immediately verifies that $u$ and thus the entire series \Ref{Uexpansion} is indeed
invariant under (\ref{diagiso}) and thus under the full diagonal $SO(4)^{(D)}$.

{}From these scalar invariants we can construct the invariant vector harmonics as
 \begin{equation}
  \begin{split}
   {\mathcal X}_a &= e_{a}{}^{m}\,\partial_m u =
   e_{a}{}^{m}\,\Big(y_{\bar m} -    \sqrt{\frac{1-y^2}{1-x^2}}\hspace{0.1em}x_m\Big)\;, \qquad
   {\mathcal X}_{\bar{a}} = 0\;, \\
   {\mathcal Y}_{\bar{a}} &=
   e_{\bar a}{}^{\bar m}\,\partial_{\bar{m}}u =
   e_{\bar a}{}^{\bar m}\,\Big(
   x_m - \sqrt{\frac{1-x^2}{1-y^2}}\hspace{0.1em}y_{\bar m}\Big)\;, \qquad
   {\mathcal Y}_a = 0\;,
  \end{split}
\end{equation}
in flat indices $a, \bar{a}$ on $S^{3}\times S^{3}$.
Under (\ref{diagiso}) they transform as under a Lorentz transformation.
We note the useful relations
${\mathcal X}_a{\mathcal X}^a={\mathcal Y}_{\bar{a}} {\mathcal Y}^{\bar{a}} = 1-u^{2}$.

The invariant tensor harmonics ${\mathcal Z}_{a\bar{b}}$
can be constructed along the same lines. 
Begin with a bivector $(1,1)$.
It follows from~(\ref{ExpansionVector})
(and its analogue on the second $S^{3}$) that invariant tensors in the
Kaluza-Klein tower on top of a bivector $(1,1)$ can arise from either of the series
of representations
$[j,j;j,j]$, $[j\!+\!1,j\!+\!1;j,j]$, and $[j,j;j\!+\!1,j\!+\!1]$.
Indeed, there are three independent
tensor harmonics ${\mathcal Z}^0_{a\bar{b}}$, $\CZ_{a\bar{b}}^\pm$
which can explicitly be constructed as
 \bea
  \CZ^0_{a\bar{b}}&\equiv&e_a^{\hspace{0.2em}m}\partial_m \mathcal{Y}_{\bar{b}}\
  = e_{\bar{b}}^{\hspace{0.2em}\bar{m}}\partial_{\bar{m}} \mathcal{X}_a\;,
 \qquad
   \CZ_{a\bar{b}}^\pm
   ~\equiv~ u \CZ^{0}_{a\bar{b}}-{\cal X}_a {\cal Y}_{\bar{b}}\pm
   \epsilon_a^{\hspace{0.5em}cd}{\cal X}_c \CZ_{d\bar{b}}
   \;.
 \eea
The most general deformation of the metric~(\ref{sphere}) on $S^{3}\times S^{3}$
preserving the diagonal isometries~\Ref{diagiso} can then be described by
the six-dimensional vielbein (in triangular gauge)
 \bea\label{defbein}
   E_m^{\hspace{0.3em}a} &=& g\,e_m^{\hspace{0.3em}b}\left(
   a(u)\delta_b^{\hspace{0.3em}a} + c_1(u){\cal X}_b {\cal X}^a\right)\;,
   \qquad
   E_{\bar{m}}^{\hspace{0.3em}\bar{a}} = g\,e_{\bar{m}}^{\hspace{0.3em}\bar{b}}
   \left( b(u)\delta_{\bar{b}}^{\hspace{0.3em}\bar{a}}
   + c_2(u){\cal Y}_{\bar{b}} {\cal Y}^{\bar{a}}\right)\;,  \nonumber\\
   E_m^{\hspace{0.3em}\bar{a}}
   &=& g\,e_m^{\hspace{0.3em}b}\left(d(u) \CZ^{0\hspace{0.1em}\bar{a}}_b
   +d_+(u) \CZ_b^{+\hspace{0.1em}\bar{a}}+d_-(u) \CZ_b^{-\hspace{0.1em}\bar{a}}
   \right)\;,
   \qquad E_{\bar{m}}^{\hspace{0.3em}a}=0\;,
 \eea
with a priori seven undetermined functions of $u$,
and the constant $g$ related to the sphere radius as $g=R$.
By fixing part of the diffeomorphism symmetry some of the free functions
can be set to zero. Namely, employing a diffeomorphism
 \bea\label{genvec}
  \xi^m = f(u){\cal X}^m\;,\qquad
  \xi^{\bar m} = g(u){\cal Y}^{\bar m}\;,
 \eea
the functions
$f(u)$ and $g(u)$ can be chosen such that $c_{1}(u)=c_{2}(u)=0$ in~(\ref{defbein}).

Similarly the most general ansatz for the 3-form flux
compatible with  the diagonal isometries can be constructed.
To this end, we write
 \bea\label{backflux}
&&  F_{mnk}=
  \kappa\,\omega_{mnk}+3\partial_{[m}c_{nk]}
  \;, \quad
  F_{\bar{m}\bar{n}\bar{k}}=
  \kappa\,\omega_{\bar{m}\bar{n}\bar{k}}+3\partial_{[\bar m}c_{\bar n\bar k]}
  \;,\nonumber\\
 && F_{mn\bar k}=3\partial_{[m}c_{n\bar k]}
 \;,\quad
 {\rm etc.}\;,
 \label{3formdef}
 \eea
 with
 \bea\label{Bfluc}
   c_{mn} &=& b_1(u)\omega_{mnk}{\cal X}^k\;, \qquad
   c_{\bar{m}\bar{n}} = b_2(u)\omega_{\bar{m}\bar{n}\bar{k}}
   {\cal Y}^{\bar{k}}\;, \\
   c_{m\bar{n}} &=& b_3(u){\mathcal Z}_{m\bar{n}} + b_+(u) {\mathcal Z}_{m\bar{n}}^+
   +b_-(u){\mathcal Z}_{m\bar{n}}^-\;.
 \eea
Here $\kappa=2g^{2}$ and
$\omega_{mnk}$ and $\omega_{\bar{m}\bar{n}\bar{k}}$ denote the volume forms
on the undeformed spheres $S^{3}$, respectively.
The tensor gauge symmetry can be used
to set the component $b_3(u)$ to zero.

With the most general ansatz compatible with the symmetry~\Ref{isodef}
at hand, we can now solve the IIB field equations \Ref{IIBeq},
with the deformed flux satisfying \Ref{Hconst}.
This leads to a highly
complicated nonlinear system of differential equations for the functions
$a(u)$, $b(u)$, $d(u)$, $d_{\pm}(u)$, $b_{1,2}(u)$, $b_{\pm}(u)$.
Rather than attempting a solution in closed form we expand the system
in the deformation parameter $\rho$ and solve it order by order in $\rho$
(using {\tt Mathematica}).
Further imposing antisymmetry of $\rho$ under exchange of the
two spheres, we find for the metric
\bea
   a(u)&=&
   1+ u\rho + \ft12 u^2 \rho^2+ \ft12 u^3 \rho^3 + {\cal O}(\rho^4)\;, \\[1ex]
   \quad b(u)&=&
    1- u\rho + \ft12 u^2 \rho^2- \ft12 u^3 \rho^3 +{\cal O}(\rho^4)\;, \\[1ex]
   d(u)&=&-2u\rho^2\,(1+ u\rho) +O(\rho^4)  \;, \qquad   d_\pm(u)={\cal O}(\rho^4)\;,
 \eea
while for the 3-form solution we obtain
 \bea
   b_1(u)&=&-2\rho\,\Big(
   1-u\rho+(u^{2}\!+\!\ft23)\,\rho^{2}\Big) + {\cal O}(\rho^4)
\;, \\
   b_2(u)&=& 2\rho\,\Big(
   1+u\rho+(u^{2}\!+\!\ft23)\,\rho^{2}\Big) + {\cal O}(\rho^4) \;,\qquad
   b_\pm(u)=\pm\ft43\,u\rho^{3}+{\cal O}(\rho^4)\;. 
 \eea
In particular, we see that to lowest order in $\rho$ the
deformation just corresponds to a relative warping between the two spheres.
At higher orders, also off-diagonal components of the metric are
excited.

\section{The Kaluza-Klein spectrum}
\label{sec:KKspectrum}

Before discussing the Kaluza-Klein spectrum of fluctuations around
the deformed near-horizon limit constructed in the previous
section, we first have to review the spectrum on the undeformed
background~\Ref{AdSSSS}. Its isometry supergroup under which the
spectrum is organized is the direct product of two ${\cal N}=4$
supergroups \bea D^1(2,1;\alpha)_{L} \times D^1(2,1;\alpha)_{R}
\;, \label{supergroup} \eea in which each factor combines a
bosonic $SO(3)\times SO(3)\times SL(2,\mathbb{R})$ with eight real
supercharges~(see~\cite{Sevrin:1988ew} for definitions).
More precisely, the noncompact factors $SL(2,\mathbb{R})_{L}\times
SL(2,\mathbb{R})_{R} = SO(2,2)$ join into the isometry group of
$AdS_3$ while the compact factors build up the isometry groups
$SO(4)\times SO(4)'$ of the two spheres. The parameter $\alpha$
of~(\ref{supergroup}) describes the ratio of the radii of the two
spheres $S^3$, i.e.~the ratio of D5 brane charges, which we have set to one. We note that
$D^1(2,1;1)=O\!Sp(4|2,\mathbb{R})$.

The massive Kaluza-Klein spectrum of maximal nine-dimensional supergravity on
the $AdS_3\times S^3\times S^3$ background has been computed in
\cite{deBoer:1999rh}. We give a short review of the computation in Appendix~\ref{A}.
The resulting three-dimensional spectrum can be summarized as
 \begin{eqnarray}\label{tower}
&&
\bigoplus_{\ell\ge0,\ell'\ge1/2} (\ell,\ell';\ell,\ell')_{\rm S}~~\oplus
   \bigoplus_{\ell\ge 1/2,\ell'\ge 0}(\ell,\ell';\ell,\ell')_{\rm S} \nonumber\\[1ex]
&&
\qquad\qquad
   \oplus\bigoplus_{\ell,\ell'\ge 0}\big( (\ell,\ell';\ell\!+\ft12,\ell'\!+\ft12)_{\rm S}
   \oplus (\ell\!+\ft12,\ell'\!+\ft12;\ell,\ell')_{\rm S} \big)
   \;,
 \end{eqnarray}
in terms of supermultiplets built from left-right tensor products of the
short supermultiplets $(\ell,\ell')_{\rm S}$ of $O\!Sp(4|2,\mathbb{R})$~\cite{Gunaydin:1986fe},
summarized in table~\ref{shortD}.
Note that the resulting multiplets $(\ell,\ell';\ell,\ell')_{\rm S}$ generically
contain massive fields with spin running from 0 to $\ft32$, whereas
multiplets of the type $(\ell,\ell';\ell\!+\!\ft12,\ell'\!+\!\ft12)_{\rm S}$
represent massive spin-2 multiplets.
The lowest massive multiplets in the spectrum~\Ref{tower}
are somewhat degenerate and collected in Table~\ref{spinspin};
will refer to these as the spin-$\frac12$ matter multiplet and the
(massive) YM multiplet, respectively.

\begin{table}[bt]
\centering
  \begin{tabular}{c | c c c}
  $h$ & \\
  \hline
   $h_{0}$ & & $(\ell,\ell')$ & \\[.5ex]
   $h_{0}+\frac{1}{2}$ & $(\ell-\frac{1}{2},\ell' -\frac{1}{2})$ &
   $(\ell +\ft12,\ell' -\ft12)$ & $(\ell -\ft12,\ell' +\ft12)$ \\[.5ex]
   $h_{0}+1$ & $(\ell,\ell' -1)$ &
   $(\ell -1,\ell')$ & $(\ell,\ell')$ \\[.5ex]
   $h_{0}+\ft32$ & & $(\ell -\ft12,\ell' -\ft12)$ &
  \end{tabular}
  \caption{\small The generic
  short supermultiplet $(\ell,\ell')_{\rm S}$ of $O\!Sp(4|2,\mathbb{R})$,
with  $h_{0}=\frac12(\ell+\ell')$.}
  \label{shortD}
\end{table}

\begin{table}[bt]
\centering
  \begin{tabular}{c||c|c|}
   \raisebox{-1.25ex}{$h_L$} \raisebox{1.25ex}{$h_R$} &
   $\frac{1}{4}$ & $\frac{3}{4}$
   \rule[-2ex]{0pt}{5.5ex}\\
    \hline\hline
   $\frac{1}{4}$ & $(0,\ft12;0,\ft12)$ & $(0,\ft12;\ft12,0)$
   \rule[-1.5ex]{0pt}{4ex}\\
    \hline
   $\frac{3}{4}$ &  $(\ft12,0;0,\ft12)$ &
   $(\ft12,0;\ft12,0)$  \rule[-1.5ex]{0pt}{4ex} \\
   \hline
  \multicolumn{3}{c}{\vphantom{I}}\\[.9ex]
  \end{tabular}
\qquad\;\;
  \begin{tabular}{c||c|c|c|}
   \raisebox{-1.25ex}{$h_L$} \raisebox{1.25ex}{$h_R$} &
   $\frac12$ & $1$
   & $\frac32$
   \rule[-2ex]{0pt}{5.5ex}\\
    \hline\hline
   $\frac12$ & $(0,1;0,1)$  & $(0,1;\ft12,\ft12)$ & $(0,1;0,0)$
   \rule[-1.5ex]{0pt}{4ex}\\
    \hline
   $1$ &  $(\ft12,\ft12;0,1)$
   & $(\ft12,\ft12;\ft12,\ft12)$ &
   $(\ft12,\ft12;0,0)$  \rule[-1.5ex]{0pt}{4ex} \\
     \hline
   $\frac32$ &  $(0,0;0,1)$ & $(0,0;\ft12,\ft12)$
   & $(0,0,0,0)$ \rule[-1.5ex]{0pt}{4ex} \\
  \hline
  \end{tabular}
      \caption{\small The spin-$\ft12$ multiplet $(0,\ft12;0,\ft12)_{\rm S}$,
      and the spin-$1$ multiplet~$(0,1;0,1)_{\rm S}$.}
\label{spinspin}
\end{table}

Included in~\Ref{tower} is the massless supergravity multiplet
$(\ft12,\ft12;0,0)_{\rm S}\oplus (0,0;\ft12,\ft12)_{\rm S}$
which contains no propagating degrees of freedom and
consists of the vielbein, eight gravitinos transforming as
\bea
\psi_{\mu}^{I} &:&\;\;
 (\ft12,\ft12;0,0)\oplus (0,0;\ft12,\ft12) \;
 \label{supercharges}
\eea
under (\ref{iso}), and topological gauge vectors, corresponding to the
$SO(4)_{L}\times SO(4)_{R}$ gauge group
of the effective three-dimensional theory.
The effective three-dimensional theories describing the coupling of
the supergravity multiplet to the lowest massive supermultiplets from~\Ref{tower}
have been constructed in~\cite{Nicolai:2001ac,Nicolai:2003ux,Hohm:2005ui}.

In order to study the deformation of the spectrum and the associated effective theory,
we need to identify the
field corresponding to the deformation parameter $\rho$ within~\Ref{tower}.
Since the deformation breaks
supersymmetry ${\cal N}=(4,4)\rightarrow{\cal N}=(3,3)$ and
the isometry group $SO(4)\times SO(4)'$ down to
the diagonal~\Ref{isodef}, it should be contained in a scalar representation
of the type $(\ell_{L},\ell_{L};\ell_{R},\ell_{R})$, with $\ell_{R}$
and $\ell_{L}$ not both equal to zero. Moreover,
since the deformation preserves the $AdS_{3}$ factor, the corresponding
field should have no AdS mass, i.e.~come with boundary conformal
dimension $\Delta=2$.
{}From Table~\ref{scalars0} we identify four possible candidates
with $\Delta=2$: two in the
$(\frac12,\frac12;\frac12,\frac12)$ representation and sitting in
the spin-1 (YM) multiplets of Table~\ref{spinspin}, and two in
the $(1,1;1,1)$ representation that originate from higher supermultiplets.
Note that these fields come with $\Delta=2$ only for $\alpha=1$,
i.e.\ coinciding D5 brane charges, in accordance with the above remarks
about the existence of the brane reconnection.

To narrow down which of the representations 
$(\frac12,\frac12;\frac12,\frac12)$ and $(1,1;1,1)$ 
actually contain the deformation, we make use of the fact that
according to its definition~\Ref{rhoeq}
$\rho$ should be odd under exchange of the two spheres.
The two $(\frac12,\frac12;\frac12,\frac12)$ descend from chiral multiplets
($(0,1;0,1)_{{\rm S}}$ and $(1,0;1,0)_{{\rm S}}$),
so will have one odd and one even combination, whereas the
two $(1,1;1,1)$  come from nonchiral multiplets.
Thus, parity suggests 
that the only combination of fields odd under exchange of the two spheres
is a combination of the two $(\frac12,\frac12;\frac12,\frac12)$ scalars.
In order to study the $\rho$ deformation in the effective three-dimensional
theory we will thus have to consider the coupling of two YM multiplets.
This is the goal of the next section.
Indeed,  in this effective theory we find a potential with a flat direction
(Figure~\ref{fig:valley} below) for the aforementioned combination,
along which supersymmetry is broken from
${\cal N}=(4,4)$ down to ${\cal N}=(3,3)$.
We take this as strong support of our claim
that the deformation in fact arises from 
the $(\frac12,\frac12;\frac12,\frac12)$ representation
and not the $(1,1;1,1)$.

Let us close this section by a few general remarks
on ${\cal N}=(4,4)\rightarrow {\cal N}=(3,3)$
supersymmetry breaking.
In terms of supergroups this corresponds to the natural embedding
 $O\!Sp(3|2,\mathbb{R})\subset O\!Sp(4|2,\mathbb{R})$.
A short $O\!Sp(3|2,\mathbb{R})$ supermultiplet $(\ell)_{\rm S}$ is defined by its
highest weight state $(\ell)^{h_{0}}$, where $\ell$ labels the
$SO(3)$ spin and $h=h_{0}=\ell/2$ is the charge under the Cartan
subgroup ~$SO(1,1)\subset SL(2,\mathbb{R})$. The short
supermultiplet is
generated from the highest weight state by the action of two out
of the three supercharges and carries $8\ell$ degrees of freedom~\cite{Gunaydin:1987hb}.
Its $SO(3)^{\pm}$ representation content is summarized in
Table~\ref{short3}.

\begin{table}[b]
 \centering
  \begin{tabular}{c | c  }
  $h$ & \\
  \hline
   $h_{0}$ &  $(\ell)$  \\[.5ex]
   $h_{0}+\frac{1}{2}$ & $(\ell)+(\ell\!-1\!)$  \\[.5ex]
   $h_{0}+1$  & $(\ell\!-1\!)$
  \end{tabular}
  \caption{\small The generic
  short supermultiplet $(\ell)_{\rm S}$ of $O\!Sp(3|2,\mathbb{R})$,
  with  $h_{0}=\ell/2$.}
  \label{short3}
\end{table}

The generic long multiplet $(\ell)_{\rm long}$ is  instead built
from the action of all three supercharges on the highest weight
state and correspondingly carries $8(2\ell+1)$ degrees of freedom.
Its highest weight state satisfies the unitarity bound
\be
h\ge\ell/2  \; .   \label{Longn3}
\ee
Here it is worthwhile to pause and contrast the simplicity of this unitarity
bound with the {\it nonlinear} bound
for the unbroken large ${\cal N}=4$ algebra. 
This nonlinearity was responsible for many of the complications 
in constructing holographic dual pairs for the large ${\cal N}=4$ theory 
\cite{deBoer:1999rh,Gukov:2004ym}. For example, 
states that are classically BPS can receive quantum corrections
in the large ${\cal N}=4$ theory, an unusual situation.
By comparison, a bound as simple as
\Ref{Longn3} seems a compelling reason
 for studying ${\cal N}=3$ theories in further detail,
 both in their own right and for their connection with  ${\cal N}=4$
 theories. (A nice summary is given in \cite{Miki:1989ri}).

When the bound \Ref{Longn3} is saturated, the long multiplet
decomposes into two short multiplets according to
\bea
 (\ell)_{\rm long} &=&
 (\ell)_{\rm S} \oplus (\ell\!+\!1)_{\rm S} \;,
 \label{long3}
\eea
from which one may read off the $SO(3)$ content of $(\ell)_{\rm long}$.
 A semishort ${\cal N}=4$ multiplet $(\ell^+,\ell^-)_{{\rm S}}$
breaks according to
 \bea
  (\ell,\ell')_{{\rm S}} &=&
  (\ell\!+\ell')_{{\rm S}} \oplus
  (\ell\!+\ell'\!\!-\!1)_{{\rm long}} \oplus ~\dots~ \oplus
  (|\ell\!-\ell'|)_{{\rm long}}\;,
  \label{breakN43}
 \eea
into semishort and genuine long ${\cal N}=3$ multiplets.
{}From~\Ref{breakN43} one may read off the decomposition
of the spectrum~\Ref{tower} after turning on the deformation.
The masses of the long multiplets are not protected and may
acquire $\rho$-dependent deformation contributions. In principle, even
semi-short multiplets originating from different ${\cal N}=4$ multiplets
may recombine according to~\Ref{long3} into long ${\cal N}=3$ multiplets
and lift off from the mass bound along the deformation.
We will see an example of this in the next section.

\section{The effective action in $D=3$}

In this section we discuss the effective supergravity
action in three dimensions, which describes the
YM (spin-1) multiplets $(1,0;1,0)_{\rm S}\oplus (0,1;0,1)_{\rm S}$
discussed above.
In particular, we compute the scalar potential. At the origin of scalar field space we have
the undeformed background $AdS_{3}\times S^{3}\times S^{3}$.
We show that there is a flat direction along which supersymmetry is broken down
to ${\cal N}=(3,3)$ and compute the
deformation of the mass spectrum along the valley.

\subsection{Effective action for the YM multiplets}

To start with, we note that in accordance with the
amount of supersymmetry preserved by the undeformed background,
the relevant three-dimensional supergravity will be a gauged ${\cal N}=8$ theory
with gauge group $SO(4)\times SO(4)$.
Here, we briefly review the construction of the effective theory
based on~\cite{Nicolai:2001ac,Nicolai:2003ux,Hohm:2005ui}
to which we refer for details.

The field content of the two YM multiplets $(1,0;1,0)_{\rm S}\oplus (0,1;0,1)_{\rm S}$
is given in Table~\ref{spinspin}, in particular they contain 32 bosonic degrees of freedom each.
Together with ${\cal N}=8$ supersymmetry
this implies that the scalar degrees of freedom of the three-dimensional theory
are described by a coset space  $SO(8,8)/(SO(8)\times SO(8))$.
(The massive vector degrees of freedom appear through their Goldstone scalars).
This in turn requires an embedding of the gauge group $SO(4)\times SO(4)$
into $SO(8)\times SO(8)$, such that the corresponding branching of the
${\bf (8,8)}$ representation of the latter reproduces the correct $SO(4)\times SO(4)$
representations of Table~\ref{spinspin}.
The explicit embedding was given in \cite{Nicolai:2003ux,Hohm:2005ui},
and is described by a constant $SO(8,8)$ tensor $\Theta_{\cal MN}=\Theta_{\cal (MN)}$
in the symmetric product of two adjoint representations 
whose explicit form determines the entire Lagrangian.

Explicitly, the action is given by
 \bea
  {\cal L}=-\frac{1}{4}\sqrt{g}R +\frac{1}{4}\sqrt{g}
   {\cal P}_{\mu}^{Ir}{\cal P}^{\mu \hspace{0.2em}Ir}
   +{\cal L}_{\text{CS}} +{\cal L}_{\rm ferm}-\sqrt{g}V\;,
\label{action}
 \eea
 where the individual ingredients are as follows.
The scalar fields are described by an $SO(8,8)$ valued matrix and
its current
\bea
{\cal S}^{-1}(\partial_{\mu}+g\Theta_{\cal MN}A_{\mu}^{\cal M}t^{\cal N})
{\cal S} &=& \ft12{\cal Q}_{\mu}^{IJ}X^{[IJ]}
+\ft12{\cal Q}_{\mu}^{rs}X^{[rs]}+{\cal P}_{\mu}^{Ir}X^{Ir}
\;,
\eea
decomposed into compact (${\cal Q}_{\mu}$) and noncompact (${\cal P}_{\mu}$)
contributions. Indices $I,J,\dots$ and $r,s,\dots$ are
vector indices of the two $SO(8)$ subgroups;
adjoint $SO(8,8)$ indices ${\cal M,N}$ thus split into pairs
$([IJ],[rs],Ir)$.
The vector fields couple by a Chern-Simons term
\ba
\CL_{\rm CS}&=&
-\ft14\Geps^{\mu\nu\rho} A^\cM_\mu \,\GTh_{\cM\cN}\,
\left( \dd_\nu A^\cN_\rho
+ \ft13\,
f^{\cN\cP}{}_{\cL}\,\GTh_{\cP\cK}\,A^\cK_\nu A^\cL_\rho\right)
\;,
\ea
with the $\SO{8,8}$ structure constants $f^{\cN\cP}{}_{\cL}$. The potential $V$ is given as a
function of the scalar fields as
\ba
V &=&  -\ft1{48} \left(
T_{[IJ,KL]}T_{[IJ,KL]}+ \ft1{4!}\,\Ge^{IJKLMNPQ}\, T_{IJ,KL}T_{MN,PQ}
- 2\, T_{IJ,Kr}T_{IJ,Kr}
\right) \;,
\la{W}
\ea
in terms of the so-called $T$-tensor
\ba T_{IJ,KL} &=&
\CV^{\cM}{}_{\!IJ}\CV^{\cN}{}_{\!KL}\,\GTh_{\cM\cN} \;, \qquad
T_{IJ,Kr} ~=~ \CV^{\cM}{}_{\!IJ}\CV^{\cN}{}_{\!Kr}\,\GTh_{\cM\cN}
\;, \la{T} \ea
where $\CV$ defines the group matrix $\CS$ in the adjoint
representation:
\ba \CS^{-1} t^\cM \CS &\equiv&  \ft12\,
\CV^{\cM}{}_{\!IJ} \,X^{IJ}+ \ft12\, \CV^{\cM}{}_{\!rs}
\,X^{rs}+\CV^{\cM}{}_{\!Ir} \,Y^{Ir} \;. \la{V} \ea
For the fermionic contributions ${\cal L}_{\rm ferm}$
we refer to \cite{Nicolai:2001ac}.

\subsection{The marginal ${\cal N}=(3,3)$ deformation}

We are mainly interested in the scalar potential~\Ref{W}. A
$\rho$-dependent deformation that preserves $AdS_{3}$ while
deforming the two spheres and breaking the symmetry according
to~\Ref{isodef} should manifest itself in the existence of a
corresponding flat direction of the potential. The full
potential~\Ref{W}, being a rather complicated function of the 64
scalar fields, is not needed. For our purposes it will be
sufficient to consider its truncation to $SO(4)^{(D)}$ 
(defined in \Ref{isodef}) singlets.
Indeed, extremal points in this truncated potential will lift to
extremal points of the full potential~\cite{Warner:1983vz}.

Under  $SO(4)^{(D)}$, each YM multiplet contains two scalar singlets,
i.e.\ we have a four-dimensional manifold of scalars invariant under
$SO(3)^{(D)}_L\times SO(3)^{(D)}_R$. At the origin, these scalars come
in two pairs with square masses $0$ and $3$, i.e.~they correspond to
operators of conformal dimensions $\Delta=2$ and $\Delta=3$.
In particular, there are two marginal operators,
in accordance with Table~\ref{spinspin}, above.
In order to describe the truncation of the Lagrangian
to this four-dimensional target space manifold, we
parametrize the $SO(8,8)$ matrix ${\cal S}$ as
\bea
{\cal S}&=& \exp \left(
\begin{array}{cccc}
0&0&v_{1}&w_{2}\\
0&0&w_{1}&v_{2}\\
v_{1}&w_{1}&0&0\\
w_{2}&v_{2}&0&0
\end{array}\right)
\;,
\label{4dsubspace}
\eea
where each entry represents a multiple of the
$4\times4$ unit matrix.
Note that $v_{1}$, $v_{2}$ parametrize the two $SO(4)\times SO(4)'$
singlets.
(Truncation to a single YM multiplet would correspond to setting
$w_{1}=v_{2}=0$.)
Unfortunately, even the truncation of the potential ~\Ref{W} to the four-dimensional
subspace \Ref{4dsubspace} is a highly complicated function. We 
computed it using {\tt Mathematica} but refrain from giving it here.~\footnote{It can be found at
{\tt www.aei.mpg.de/$\sim$mberg/physics/potential}.}
Instead, we further truncate to the two-dimensional
subspace defined by $v_{1}=v_{2}$, $w_{1}=-w_{2}$.
This again is a consistent truncation as it corresponds to
the fixed points of an inner automorphism that leaves $\Theta_{\cM\cN}$
invariant.

\begin{figure}[t]
\vspace{-5mm}
\begin{center}
\psfrag{t}[bc][bc][1][0]{$\tau$}
\includegraphics[width=0.6\textwidth]{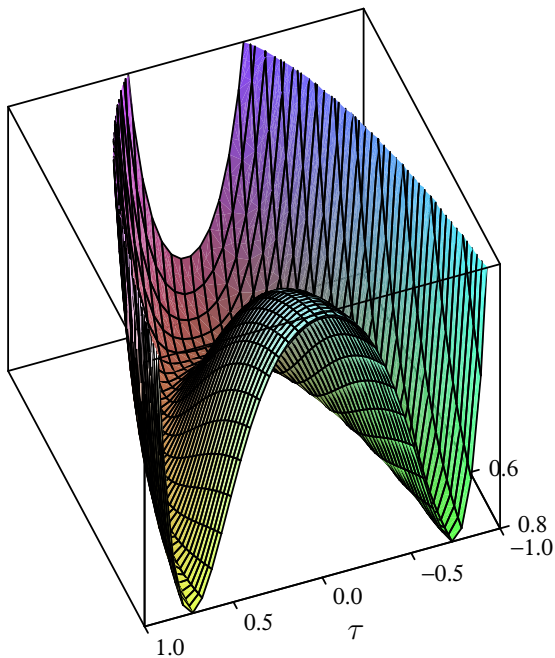}
\vspace{-5mm}
\caption{Flat valley in the potential, the ${\cal N}=(4,4)$
origin is located at $(0,0)$. Coordinates here are
$(\tau,\zeta)$ as in \Ref{xycoords}.}
\label{fig:valley}
\end{center}
\vspace{-5mm}
\end{figure}

In terms of the variables
\bea
z^{2}=v_{1}^{2}+w_{1}^{2}\;, \quad
\phi=\arctan (w_{1}/v_{1})\;,
\eea
this gives rise to a Lagrangian
\bea
{1 \over \sqrt{g}}{\cal L}=
\partial_{\mu}z\,\partial^{\mu}z+
\sinh^{2}\!z\,\partial_{\mu}\phi\,\partial^{\mu}\phi
-V
\;,
\eea
with the scalar potential
\bea
V&=&-2+
8\sinh^2\!z \,
(\sinh\!z-\cos\phi \cosh\!z)^2\,
(1 + 2\cosh2z- 2\cos\phi\sinh2z)
 \;.
 \label{Vtrunc}
\eea
Obviously, this potential is bounded from below ($V\ge-2$). Further transforming to
coordinates
\be
\label{xycoords}
\tau=\sin\phi\sinh z\, , \;
\zeta=\cos\phi\sinh z \; ,
\ee
we find that the minimum $V=-2$ is actually taken along a curve
\be
\zeta = {\tau \over \sqrt{1-\tau^2}}
\label{val}
\ee
with $\tau$ running from 0 to 1,
which thus constitutes a
flat direction in the potential, depicted in Figure~\ref{fig:valley}.
We have verified by explicit computation that this extends to a flat direction
in the full four-dimensional target space~\Ref{4dsubspace}
and thus of the full scalar potential.
In terms of the coordinates $\tau$, $\zeta$,
exchange of the two spheres corresponds to $\tau \rightarrow-\tau$, so
the graph shows that infinitesimally, the valley indeed points
into an odd direction in accordance with the odd parity of 
the deformation $\rho$ discussed in earlier sections.
In other words, we can identify $\tau=\rho$ to lowest order.

By construction, any nonvanishing expectation value of the scalar
fields corresponding to~\Ref{val} yields an $AdS_{3}$ solution of
the three-dimensional theory which breaks the original
$SO(4)\times SO(4)'$ symmetry down to the diagonal
and supersymmetry down to ${\cal N}=(3,3)$ as we shall see below. 
From a purely supergravity point of view the existence of this flat direction is
 surprising, but it finds a natural interpretation in terms
of the brane reconnection described in earlier sections. Note further that in
the three-dimensional theory the deformation \Ref{val} is very simple
(once
the scalar potential~\Ref{Vtrunc} has been computed) and exact to
all orders in the deformation, whereas in 10 dimensions we have only been
able to perturbatively compute the corresponding solution. This
shows how nontrivially the effective theory~\Ref{action} must be
embedded within the IIB theory. Techniques such as those employed
in~\cite{Pilch:2000fu,Pilch:2003jg} might prove useful to obtain
the corresponding IIB solution in closed form.

\subsection{Deformed spectrum}

Having established a one-parameter class of solutions
of the three-dimensional effective theory, we can now study
how the mass spectrum changes upon moving along the valley \Ref{val},
that we can parameterize by $\tau$. As $\tau$ is odd under exchange of the two spheres,
the spectrum should be even in $\tau$.

As the deformation preserves the diagonal subgroup~\Ref{isodef},
the deformed spectrum organizes under~$SO(4)^{(D)}$.
Of particular interest is the remaining supersymmetry.
This is found by calculating the deformation of the
gravitino masses as eigenvalues of~\cite{Nicolai:2001ac}
 \bea
  A_1^{AB}&=&-\delta^{AB}\theta-\frac{1}{48}\Gamma_{AB}^{IJKL}T_{IJ|KL}\;.
 \eea
As a result we find
\bea
m_{i}=\pm\ft12\quad (\times 3)\;,
\qquad m_{i}=\pm\ft12f(\tau)\quad (\times 1)\;,
\eea
where $f(\tau)$ is given in terms of the coordinate $\tau$ in Fig.~\ref{fig:valley}
by
\be
f(\tau) = \sqrt{1+15\tau^2 \over 1-\tau^2}
\ee
Thus, we see that
when we move away from the origin along the valley $(\tau>0)$,
 supersymmetry is broken from ${\cal N}=(4,4)$ down
to ${\cal N}=(3,3)$, confirming the arguments in earlier sections.
From the point of view of the 3-dimensional effective theory,
this was by no means guaranteed.

By linearizing the full scalar potential~\Ref{W} around the deformed solution,
we obtain the deformed scalar masses, most conveniently expressed
in terms of the associated conformal dimensions $\Delta=1+\sqrt{1+m^{2}}$:
\bea
\Delta_{i}&=&\left\{
\begin{array}{ll}
1 & (\times  9)\\
2 & (\times 34)\\
3 & (\times 1)\\
\ft12\,(-1+f(\tau)) & (\times 1)\\
\ft12\,(1+f(\tau)) & (\times 9)\\
\ft12\,(3+f(\tau)) & (\times 9)\\
\ft12\,(5+f(\tau)) & (\times 1)
\end{array}\right.
\label{scalardeltas}
\eea
{}From these values we can infer the entire spectrum
in terms of ${\cal N}=(3,3)$ supermultiplets.
Comparing~(\ref{scalardeltas}) to table~\ref{short3}
we conclude that the ${\cal N}=(3,3)$ spectrum along the deformation
is given by
\bea
{\cal H}_{\rho}&=&
(1;1)_{\rm S} \oplus (0;0)^{h}_{\rm long} \;,
\label{N3spectrum}
\eea
where the mass of the long multiplet is given by $h=\ft14\,(-1+f(\tau))$.
We see that the entire spectrum is indeed even in $\tau$. 

For convenience, we have collected the field content of these two multiplets
in Tables~\ref{N3}, \ref{N3long}.
As $\tau$ tends to one,
the long multiplet $(0;0)^{h}_{\rm long}$ becomes infinitely massive,
and we are left with the semi-short multiplet $(1;1)_{\rm S}$ whose coupling
to ${\cal N}=(3,3)$ supergravity is described by a gauged theory
with target space $SU(4,4)/S(U(4)\times U(4))$.
Comparing this multiplet to the original field content (Table~\ref{spinspin})
one recognizes a diagonal combination of the two ${\cal N}=4$ YM multiplets.

\begin{table}[bt]
\centering
  \begin{tabular}{c||c|c|c|}
   \raisebox{-1.25ex}{$h_L$} \raisebox{1.25ex}{$h_R$} &
   $\frac{1}{2}$ & $1$
   & $\frac{3}{2}$
   \rule[-2ex]{0pt}{5.5ex}\\
    \hline\hline
   $\frac{1}{2}$ & $(1;1)$  & $(1;0)+(1;1)$ & $(1;0)$
   \rule[-1.5ex]{0pt}{4ex}\\
    \hline
   $1$ &  $(0;1)+(1;1)$
   & $(0;0)+(0;1)+(1;0)+(1;1)$ &
   $(0;0)+(1;0)$  \rule[-1.5ex]{0pt}{4ex} \\
     \hline
   $\frac{3}{2}$ &  $(0;1)$ & $(0;0)+(0;1)$
   & $(0;0)$ \rule[-1.5ex]{0pt}{4ex} \\
  \hline
  \end{tabular}
      \caption{\small The short ${\cal N}=(3,3)$ multiplet $(1;1)_{\rm S}$.}
\label{N3}
\end{table}
\begin{table}[bt]
\centering
  \begin{tabular}{c||c|c|c|c|}
   \raisebox{-1.25ex}{$h_L$} \raisebox{1.25ex}{$h_R$} &
   $h$ & $h+\frac12$ & $h+1$ & $h+\frac{3}{2}$
   \rule[-2ex]{0pt}{5.5ex}\\
    \hline\hline
   $h$ & $(0;0)$  & $(0;1)$ & $(0;1)$ & $(0;0)$
   \rule[-1.5ex]{0pt}{4ex}\\
    \hline
   $h+\frac{1}{2}$ & $(1;0)$  & $(1;1)$ & $(1;1)$ & $(1;0)$
   \rule[-1.5ex]{0pt}{4ex}\\
    \hline
   $h+1$ & $(1;0)$  & $(1;1)$ & $(1;1)$ & $(1;0)$
   \rule[-1.5ex]{0pt}{4ex}\\
    \hline
   $h+\frac{3}{2}$ & $(0;0)$  & $(0;1)$ & $(0;1)$ & $(0;0)$
   \rule[-1.5ex]{0pt}{4ex}\\
  \hline
  \end{tabular}
      \caption{\small The long ${\cal N}=(3,3)$ multiplet $(0;0)^{h}_{\rm long}$.}
\label{N3long}
\end{table}

Now that we have the deformed spectrum, it is instructive to turn around and
study the behavior of \Ref{N3spectrum}
as the deformation is switched off ($\tau\rightarrow0$).
At this point, the long ${\cal N}=(3,3)$ multiplet hits
the unitarity bound $h=0$ and falls apart according to~\Ref{breakN43}:
\bea
(0;0)^{0}_{\rm long} &\rightarrow& (1;1)_{\rm S}\oplus
(1;0)_{\rm S} \oplus(0;1)_{\rm S} \oplus (0;0)_{\rm S} \;.
\label{gaugedofs}
\eea
Simple counting of states shows that for these low spins
the formulae degenerate such that
$(0;0)_{\rm S}$, $(1;0)_{\rm S}$, and $(0;1)_{\rm S}$ denote
unphysical multiplets without propagating degrees of freedom,
given in Table~\ref{N3neg}.
The negative multiplicities should be understood as
(first order differential) constraints that eliminate the physical degrees of freedom.
To understand the role of these unphysical multiplets at $\tau=0$ 
we have to also consider the (non-propagating) supergravity multiplet.
Applying~\Ref{breakN43} to the (unphysical) ${\cal N}=(4,4)$ supergravity multiplet
$(\ft12,\ft12;0,0)_{\rm S}\oplus(0,0;\ft12,\ft12)_{\rm S}$
shows that under ${\cal N}=(3,3)$ it decomposes as
\bea
(\ft12,\ft12;0,0)_{\rm S}\oplus(0,0;\ft12,\ft12)_{\rm S}
&\rightarrow&
(0_{\rm long};0_{\rm S})\oplus(0_{\rm S};0_{\rm long})\oplus
(1;0)_{\rm S}\oplus(0;1)_{\rm S}\;,
\eea
where the first two terms represent the ${\cal N}=(3,3)$
supergravity multiplet and in the second two terms one recognizes
the unphysical part of~\Ref{gaugedofs}.
Put together, at $\tau=0$ the long ${\cal N}=(3,3)$ multiplet
splits according to~\Ref{gaugedofs},
of which the first term
coincides with an ${\cal N}=(4,4)$ YM multiplet (Tables~\ref{spinspin},\ref{N3}),
whereas the unphysical multiplets $(1;0)_{\rm S}\oplus(0;1)_{\rm S}$ combine
with the supergravity multiplet in order to reconstitute the ${\cal N}=(4,4)$
supergravity multiplet.

Having understood how things combine when we switch
{\it off} the deformation $\tau$, we can now go back to the ${\cal N}=4$ theory
and summarize in ${\cal N}=3$ language what happens when we switch on the deformation.
Then, ${\cal N}=3$ semi-short multiplets originating from different
${\cal N}=4$ ancestors (gravity and YM multiplet)
combine to form a long ${\cal N}=3$ multiplet and lift off the mass bound.
\footnote{An analogous situation is encountered in the $AdS_{5}/{\rm CFT}_{4}$
correspondence upon switching on the 't~Hooft coupling $\lambda$.
This breaks the higher spin symmetry present at $\lambda=0$
down to $PSU(2,2|4)$. In the process, semi-short multiplets originating from different
higher-spin multiplets then combine into long multiplets of $PSU(2,2|4)$~\cite{Bianchi:2003wx}.
}

\begin{table}
  \begin{tabular}{c||c|c|c|}
   \raisebox{-1.25ex}{$h_L$} \raisebox{1.25ex}{$h_R$} &
   $0$ & $\frac12$ & $1$
   \rule[-2ex]{0pt}{5.5ex}\\
    \hline\hline
   $0$ & $(0;0)$  &  & $-(0;0)$
   \rule[-1.5ex]{0pt}{4ex}\\
    \hline
   $\frac{1}{2}$ &  & &
   \rule[-1.5ex]{0pt}{4ex}\\
    \hline
   $1$ & $-(0;0)$  &  &  $(0;0)$
   \rule[-1.5ex]{0pt}{4ex}\\
    \hline
  \end{tabular}
  \qquad\quad
    \begin{tabular}{c||c|c|c|}
   \raisebox{-1.25ex}{$h_L$} \raisebox{1.25ex}{$h_R$} &
   $0$ & $\frac12$ & $1$
   \rule[-2ex]{0pt}{5.5ex}\\
    \hline\hline
   $\frac{1}{2}$ & $(1;0)$  & & $-(1;0)$
   \rule[-1.5ex]{0pt}{4ex}\\
    \hline
   $1$ & $(0;0)+(1;0)$  & &  $-(0;0)-(1;0)$
   \rule[-1.5ex]{0pt}{4ex}\\
    \hline
   $\frac{3}{2}$ & $(0;0)$  &  & $-(0;0)$
   \rule[-1.5ex]{0pt}{4ex}\\
  \hline
  \end{tabular}
  \caption{\small Unphysical ${\cal N}=(3,3)$ multiplets $(0;0)_{\rm S}$,  $(1;0)_{\rm S}$.}
\label{N3neg}
\end{table}

\section{D1-brane CFT}

Using the holographic correspondence, we should be able to compare the spectrum
\Ref{scalardeltas} computed from supergravity to those in the CFT on the D-brane intersection.
Although a full comparison is beyond 
the scope of this paper, we give some initial steps towards this general goal.

First, we consider the symmetric product CFT $Sym^N({\bf S}^1 \times
{\bf S}^3)$,
which consists of $N$ copies of ${\bf S}^1 \times
{\bf S}^3$ orbifolded by the symmetric group $S_N$.
For equal D5-brane and D5${}'$-brane charge, this was conjectured in
\cite{deBoer:1999rh,Gukov:2004ym}  to be the CFT dual of 
Type IIB string theory on
$AdS_3 \times S^3 \times S^3 \times S^1$. 
Orbifolds of this type have been extensively studied in the literature, and
we used results from \cite{Lunin:2000yv,Balasubramanian:2005qu}.

 The bosonic part of the worldsheet action
of a D1-brane on ${\bf S}^1 \times {\bf S^3}$ is
\be
\label{ws}
S_{\rm ws} = {1 \over 2\pi} \int_{\rm D1} d^2 z\;
G_{ab}\partial X^{a}
\bar{\partial} X^b   + \int_{\rm D1} C_{(2)} \;,
\ee
where $G_{ab}$ is
the induced metric on ${\bf S}^1 \times {\bf S^3}$, $C_{(2)}$ 
is the RR 2-form potential,
and we have suppressed labels of the $N$ copies.
This is the {\it undeformed} theory. Rather than considering a sigma
model $Sym^N(\CM)$
on the full complicated $\rho$-deformed solution directly, we represent the deformation
by an operator $\CO_{\rho}$, that we obtain
by expanding the worldsheet action \Ref{ws} in
the deformation parameter $\rho$ in the full solution.
As probe of the deformation,
we then consider an  untwisted probe operator $\CO_3$, 
one of the $\Delta=3$ operators in the spectrum,
coupling to the
deformation.  We should then compute
\be
\langle \CO_3 \CO_3 \rangle_{\rho}
= \langle \CO_3 \CO_3 \rangle_{0}
+\langle \CO_3 \CO_3 \CO_{\rho} \rangle_{0} + \ldots \;,
\label{corrpert}
\ee
where the subscript $0$ refers to the undeformed theory.
Since the bulk theory has an $AdS_3$ factor also after deformation, the
boundary theory will remain conformal, and we should be able to
compute corrections to $\Delta$ this way:
\be
\Delta(\rho) = \Delta_0 + \Delta_1 \rho^2 + \ldots \;.
\ee
Since we have already computed this
deformation for all $\rho$ on the supergravity side in \Ref{scalardeltas},
we could then compare results. For now, we
will content ourselves with computing the first term in the series
\Ref{corrpert}, since for our case of Lorentzian AdS, we did not find
this done explicitly in the literature. 

\begin{figure}
\begin{center}
\psfrag{gamman1}[bc][bc][1][0]{$\gamma_n^1$}
\psfrag{gamman2}[bc][bc][1][0]{$\gamma_n^2$}
\psfrag{gamman3}[bc][bc][1][0]{$\gamma_n^3$}
\psfrag{gamman4}[bc][bc][1][0]{$\gamma_n^4$}
\psfrag{w}[bc][bc][1][0]{${w \over 2\pi}$}
\psfrag{wb}[bc][bc][1][0]{${\bar{w} \over 2\pi}$}
\psfrag{n-1}[bc][bc][0.8][0]{${n-1}$}
\psfrag{z}[bc][bc][1][0]{$z$}
\includegraphics[width=0.5\textwidth,height=6cm]{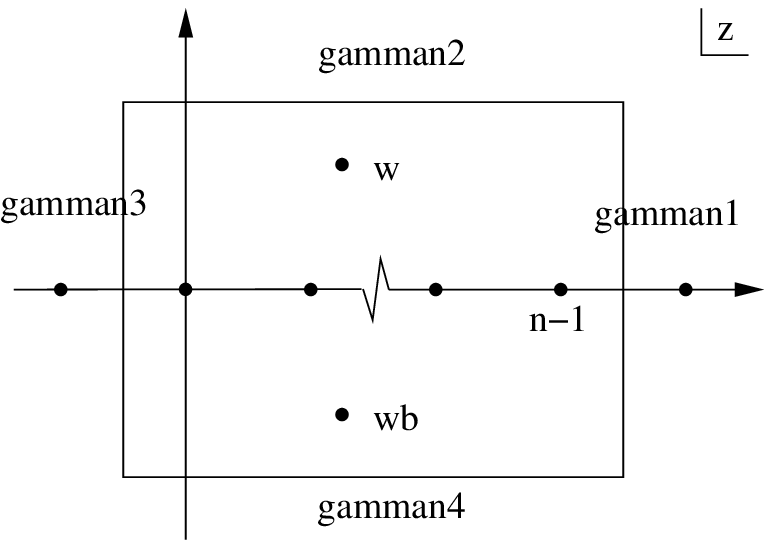}
\vspace{-2mm}
\end{center}
\caption{Jordan curve for the sum \Ref{sumk}}
 \label{fig:contour}
\end{figure}

The worldsheet coordinate is $z$.
The symmetric orbifold ground state has 
twist insertions $\sigma_{(1...n)}(z)$ at $z=0,\infty$,
and the coordinates of the $n\leq N$ copies
cyclically permute  as $X^a(z)$ encircles these points 
(see e.g.\ \cite{Lunin:2000yv}).
On the covering space $z \sim t^n$, however, all fields are
single-valued, 
so we have the ordinary 2-point function:
\be
\langle \CO_3(t_1) \CO_3(\bar{t_2}) \rangle
= {1 \over (t_1-t_2)^{3}(\bar{t}_1-\bar{t}_2)^{3}}
\;.
\ee
Going back to the $z$ variable, we obtain $n$ correlators, one
for each branch of the multiple covering,
that must be summed over. This is simpler in cylinder coordinates
$z=e^{-iw}$, where the sum is over shifts in $w$:
\be
\label{sumk}
\langle \CO_3(w_1) \CO_3(\bar{w_2}) \rangle=
\sum_{k=0}^{n-1} {1 \over (2n \sin {w-2 \pi k \over 2n})^3
(2n \sin{\bar{w}-2 \pi k \over 2n})^3}
\;,
\ee
where $w=w_1-w_2$.
This finite sum
can be performed by contour integration. To be precise,
it is performed by integrating
an analytic function $f(z)$ that has poles at $z=0, 1, \ldots, n-1$
(and possibly elsewhere)
around a suitable contour. We pick the analytic function
\be
f(z) = {\pi \cot \pi z \over (2n \sin {w-2 \pi z \over 2n})^3
(2n \sin{\bar{w}-2 \pi z \over 2n})^3}
\;,
\ee
and choose a square contour as in fig.\
\ref{fig:contour},
where the integrals over $\gamma^1_n$ and
$\gamma_n^3$ cancel by periodicity under $z\rightarrow z+n$,
and the  $\gamma^2_n$ and $\gamma^4_n$ can be moved
off to infinity, where $|f(z)|$ vanishes exponentially. Hence
the sum is the negative of the contribution from the two remaining
poles, which yields
\ba
\langle \CO_3(w_1) \CO_3(\bar{w_2}) \rangle&=&
-{1 \over 64 n^3 \sin^3 {w-\bar{w} \over 2n}}
\left[ \left( {\cos{w \over 2} \over \sin^3 {w \over 2}}
-{\cos{\bar{w} \over 2} \over \sin^3 {\bar{w} \over 2}} \right)\right.
\nonumber \\[2mm]
&&\hspace{-4cm} \left.
-{3 \over n \cot {w-\bar{w} \over 2n}}\left({1 \over \sin^2{w \over 2}}
-{1 \over \sin^2{\bar{w} \over 2}}\right)
-{6 \sin {w-\bar{w} \over 2} \over n^2 \sin^2 {w-\bar{w} \over 2n}
\sin {w \over 2} \sin{\bar{w} \over 2}}
+{4 \over n^2}{ \sin {w-\bar{w} \over 2} \over \sin {w \over 2} \sin {\bar{w}
 \over 2} }\right]
\;.
\ea
We note that all terms in the bracket except for the last persist
in the large $n$ limit, which yields\footnote{Unlike in
  \cite{Balasubramanian:2005qu}, the correlator in the large-$n$ theory
seems to have a higher-order pole than that in the covering space.}
\ba
\langle \CO_3(w_1) \CO_3(\bar{w_2}) \rangle&=&
-{1 \over 8 (w-\bar{w})^3}
\left[ \left( {\cos{w \over 2} \over \sin^3 {w \over 2}}
-{\cos{\bar{w} \over 2} \over \sin^3 {\bar{w} \over 2}} \right)
 \right.  \nonumber \\
&&  \hspace{-1cm} \left.
-6(w-\bar{w})\left({1 \over \sin^2{w \over 2}}-{1 \over \sin^2{\bar{w} \over 2}}\right)
-{24 \sin {w-\bar{w} \over 2} \over (w-\bar{w})^2
\sin {w \over 2} \sin{\bar{w} \over 2}}
\right] \; .
\ea
The task to compute the first deformed correlator is clearly more
formidable, and we will not perform it here.

\subsection{Probe approximation}
For now, a less ambitious computation would be to probe the $\rho$-deformed background 
by a ``long'' 
D1-brane probe in $AdS_3$, and quantize it semiclassically
along the lines of how it is done in $AdS_5$ (e.g.\
\cite{Frolov:2002av}, that considers both static and conformal gauge).  
In fact, for massive fluctuations in the warped region, 
we could use the simpler ``effective string
wavefunction'' argument, as for instance in \cite{Jackson:2004zg}.
We would not expect to be able to reproduce
the full contributions to the spectrum this way, of course.
For the related NS5-brane configuration in \cite{Itzhaki:2005tu},
D1-brane probes
have been studied in \cite{Kluson:2005eb,Kluson:2005qq,Kluson:2006wa}.

For this purpose, we can consider the bosonic part of the D-brane action
for a D1-brane probe in the $\rho$-deformed background in
Sec.~\ref{sec:NH}, 
with open-string vectors turned off:
 \bea
  S_{\rm D1} = {1 \over {2\pi}}\int_{\rm D1} d^2\sigma \sqrt{-\det(h_{ab})}
\; + \int_{\rm D1} C_{(2)} \;,
  \label{NGact} 
 \eea
where $h_{ab}$ is the 2-dimensional metric induced by our $\rho$-deformed
$AdS_3\times S^3\times S^3$ background:
 \bea\label{ind}
  h_{ab} = g^{AdS}_{\mu\nu}\partial_a x^{\mu}\partial_b x^{\nu}
  +  G_{ij}(X)\partial_a X^i \partial_b X^j
\;.
 \eea
Here $i,j = 1,...,6$ label the $S^3\times S^3$ coordinates,
collectively denoted by $X^i$, and $g^{AdS}_{\mu\nu}$ is the
$AdS_3$ metric. We consider a static
configuration $x^0=\sigma^0$, $x^1=\sigma^1$, $x^2=$ constant, and allow for
arbitrary fluctuations of $X^i$ around the origin. 
To zeroth order, there is no nontrivial static potential.
To second order in fluctuations,
the induced metric (\ref{ind}) becomes\footnote{Here 
we used $AdS_3$ coordinates
$ds^2 = (R^2/(2r^2))(-dt^2+dr^2+dz^2)$, unlike in 
\Ref{AdSSSS}.}
\bea
  h_{ab} = \frac{R^2}{2r^2}\eta_{ab} + G_{ij}(0)\partial_a X^i \partial_b
  X^j +O(X^3)
\;,
 \eea
with $\eta_{ab}$ denoting the flat 2d metric and $X^i$ now
meaning fluctuations. We define $\bar{\eta}_{ab} =
\frac{R^2}{2r^2}\eta_{ab}$. Expanding the square root of the
determinant to linear order (which is then second order in the
fluctuations), we find
 \bea
  \sqrt{-\det(h_{ab})} =
  \sqrt{-\det(\bar{\eta}_{ab})}\left(1+\frac{1}{2}\bar{\eta}^{ab}
  G_{ij}(0)\partial_a X^i \partial_b X^j\right) + O(X^3)
\;.
 \eea
Next let us evaluate $G_{ij}(0)$ explicitly, using
the $\rho$-deformed near-horizon limit in 
Sec.~\ref{sec:NH}. In the notation of that section, 
at $X^i =
(x^m,y^{\bar{m}}) =0$ we have $u =1$, ${\cal Z}_{a\bar{b}}^0={\cal
Z}_{a\bar{b}}^{\pm}= \delta_{a\bar{b}}$, while the other harmonics
${\cal X}_a$, ${\cal Y}_{\bar{a}}$, etc. vanish. The 
6-dimensional part of the target space metric then simply reduces
to
 \bea
  G_{ij}(0) = \left(\begin{array}{cc} (a^2(1)+d^2(1))\delta_{mn} &
  b(1) d(1)\delta_{m\bar{n}} \\ b(1)d(1)\delta_{\bar{m}n} & b^2(1)\delta_{\bar{m}\bar{n}} \end{array}\right)
\;.
 \eea
Thus, using the explicit
expressions for $a$, $b$ and $d$ at $u=1$ to the given order in
$\rho$, the quadratic fluctuation action is
 \bea
  S_{{\rm D1},\CO(X^2)} &=&  \frac{g^2}{4\pi}\int
  d^2\sigma
  \big[(1+2(\rho+\rho^2+\rho^3))\partial^ax^m\partial_ax^m\\
  \nonumber
  &&\qquad \qquad\! +(1-2(\rho-\rho^2+\rho^3))\partial^a y^{\bar{m}}\partial_a
  y^{\bar{m}}-4\rho^2\partial^ax^m\partial_a
  y^{\bar{m}}\big]
\;.
 \eea
Here everything is contracted with the flat 2d metric, i.e.\ we 
have used the fact that the conformal factor differing between
$\bar{\eta}$ and $\eta$ cancels in two dimensions.

We find that these fluctuations are massless. 
To see a mass term we would have to start from a non-constant background such
that the derivative term can give a quadratic background term,
which after expanding the scalar metric $G_{ij}(X)$ up to second
order in $X^i$ supports a mass term. One would like to expand in 
normal coordinates, i.e.
choose normal coordinates on our deformed $S^3\times S^3$ (with
$G_{ij}(X_0)=\delta_{ij}$) and then write
 \bea
  G_{ij}(X) = \delta_{ij} - \frac{1}{3}R_{ikjl}(X_0)\delta
  X^k\delta X^l + O(\delta X^3)
\;.
 \eea
To compute this explicitly in $\rho$ we would therefore
need an explicit non-trivial (non-constant) background solution
and the Riemann tensor in the corresponding normal coordinates,
which are different from ours. For massless fluctuations,
the ``effective string wavefunction'' argument is not applicable. 
There is clearly much left to do here, but
we leave this for future work.

\section{Conclusion}

In this paper, we initiated a detailed study of the $\rho$-deformed
D1-D5-D5$'$ system.
We computed the near-horizon limit of the
deformed brane configuration perturbatively in
the deformation parameter $\rho$.
Within the three-dimensional
effective gauged supergravity, we verified
the existence of a flat direction (valley) in the potential~(Figure~\ref{fig:valley})
that corresponds to the deformation,
and computed the deformed mass spectrum along the valley.

There appeared many new questions along the way.
The most glaring omission seems to be the construction of the complete
$\rho$-deformed brane solution, of which we have only constructed the
near-horizon limit. Given the technology that already exists for
related cases
(e.g.\cite{Gauntlett:1997pk,Gauntlett:1998vk,Gauntlett:1998kc}), one
could hope that this would be accomplished relatively soon. From that vantage point one
could easily answer geometrical questions left unanswered by our solution in
section \ref{sec:NH}, such as the details of the variable
transformation generalizing \Ref{NHtrans} to $\rho>0$.

With some more work along the lines of what we presented here, one
should be able to nail down the precise couplings between the $\rho$
deformation and the supergravity fluctuations in ten
dimensions.
This would pave the way for computing the
deformed correlator in \Ref{corrpert} in the CFT, leading to a highly nontrivial
comparison with the deformed $\Delta(\rho)$ in \Ref{scalardeltas}. If
successful --- and there are some pitfalls, when
using the boundary CFT at the orbifold point ---
 this would constitute one of the most detailed tests that
have ever been performed of the AdS/CFT correspondence. It is made
more feasible than most deformed correspondences by
the great simplification of the deformation being marginal,
ensuring that the boundary theory is conformal at all scales. (Indeed,
for RG flows, general $\Delta$ functions can only at best make approximate,
scheme-dependent (cf.\ \cite{Berg:2002hy})
sense away from conformal fixed points).
Using the techniques developed in~\cite{Bianchi:2001de,Bianchi:2001kw,Martelli:2002sp,Papadimitriou:2004rz}
for generic RG flows one might be able to extend the analysis 
to compute deformed (but conformal) higher-point correlation functions.

The ${\cal N}=3$ theory is interesting in its own right, not the least
because of the simpler BPS bound \Ref{Longn3}. The super-Higgs dynamics seems quite
rich in this case (cf.\ eq.~\Ref{gaugedofs}), and we would expect
a closer study of this dynamics could shed light on the
${\cal N}=3$ single D1-D5 system, as outlined in Section \ref{sec:brane}.

By analogy with the breaking ${\cal N}=4 \rightarrow {\cal N}=3$ in 4 dimensions
\cite{deRoo:1986yw,Wagemans:1987zy,Tsokur:1994gr,Frey:2002hf}, one
should be able to think of this as adding judiciously chosen flux.
A related topic of interest would be a study of deformations of the
Chern-Simons theory with flux discussed in \cite[Sec.\
  3]{Gukov:2004ym},
and \cite{Itzhaki:2005tu,Lin:2005nh}.

It would also be interesting to understand 
how our results fit into the bigger picture, if any, of
marginal deformations in AdS/CFT along the lines of
\cite{Lunin:2005jy,Frolov:2005ty,Gauntlett:2005jb}.

\section*{Acknowledgments}
We wish to thank D.~Berenstein, M.~Cederwall, S.~Cherkis, S.~Fredenhagen,
J.~Gauntlett, S.~Gukov, G.~Horowitz, A.~Kleinschmidt, J.~Maldacena,
D.~Marolf, J.~Polchinski, M. Shigemori and K.~Skenderis, 
for helpful and stimulating discussions and comments.
This work was supported in part by the European RTN Program MRTN-CT-2004-503369
and the DFG  grant SA 1336/1-1.
O.H. is supported by the
stichting FOM and the RTN network MRTN-CT-2004-005104.
M.B. is supported by The European Superstring Theory Network,
MRTN-CT-2004-512194, and would like to thank the KITP
in Santa Barbara for hospitality.
This research was supported in part by the National 
Science Foundation under Grant No. PHY99-07949. 

\begin{appendix}

\section{Appendix: Kaluza-Klein spectrum on $AdS_3\times S^3\times S^3$}
\setcounter{equation}{0}
\label{A}

In this appendix we give a brief review of the group-theoretical
analysis of the Kaluza-Klein spectrum on the $AdS_3\times
S^3\times S^3$ background, following~\cite{Salam:1981xd} (see
also~\cite{deBoer:1998ip,deBoer:1999rh,Morales:2002ys,Hohm:2006rj}). Starting
from maximal nine-dimensional supergravity, the physical fields
can be classified under the $SO(1,2)\times SO(3)\times SO(3)'$
subgroup of the nine-dimensional Lorentz group $SO(1,8)$ with the
different factors corresponding to the $AdS_3$ and the two sets of
$S^{3}$ coordinates, which we denote collectively by $z$, $x$, and
$y$, respectively. Labeling the corresponding representations by
$K$, $J$, and  $J'$, respectively, the fields can be expanded in
terms of $S^{3}$ sphere functions according to \bea
\Phi_{[K;J,J']}(z,x,y) &=& \sum_{L,L'} \phi_{[K;L,L']}(z)\,
X^{(L)}_{\!J}(x)\,Y^{(L')}_{\!J'}(y) \;. \label{KKdec} \eea The
sphere functions $X^{(L)}_{\!J}(x)$,\,$Y^{(L')}_{\!J'}(y)$ are
labeled by representations $L$, $L'$ of the isometry group
$SO(4)\times SO(4)$. The coefficients $\phi_{[K;L,L']}(z)$
describe the complete three-dimensional Kaluza-Klein spectrum. The
structure of the spectrum is thus encoded in the range of
representations $L$, $L'$ over which the sum~\Ref{KKdec} is taken.
This has been determined in~\cite{Salam:1981xd}: the sum
in~\Ref{KKdec} is running precisely over those representations
$L$, which contain the representations $J$ upon breaking of the
isometry groups $SO(4)$ down to the Lorentz groups $SO(3)$, and
similarly for $SO(4)'$.

\begin{table}[t]
\centering
  \begin{tabular}{c||c|c|c|c|c|c|c|c|c|c|c|c|}
   \raisebox{-1.25ex}{${\cal R}$} \;\;\;\raisebox{1.25ex}{$\Delta$} &
   $\frac{1}{2}$ & $1$  & $\frac{3}{2}$ & $2$ & $\frac{5}{2}$ & $3$ & $\frac{7}{2}$ & $4$
   & $\frac{9}{2}$ & $5$ & $\frac{11}{2}$ & $6$
   \rule[-2ex]{0pt}{5.5ex}\\
    \hline\hline
$[0,0;0,0]$ &&&& 2 &&2 &&2 &&&&
   \rule[-1.5ex]{0pt}{4ex}\\
    \hline
 $[\ft12,0;\ft12,0]$
 &
1
 &&
1
 &&
2
 &&
3
 &&
2 &&&
    \rule[-1.5ex]{0pt}{4ex}\\
    \hline
 $[1,0;1,0]$
 &&
1
 &&
 2
 &&
2
 &&
3 && 2
 &&
 \rule[-1.5ex]{0pt}{4ex} \\
    \hline
 $[\ft12,\ft12;\ft12,\ft12]$
 &&
 2
 &&
 2
 &&
 6
 &&
4
 &&
 2
 &&
\rule[-1.5ex]{0pt}{4ex} \\
      \hline
 $[1,\ft12;1,\ft12]$
 &&&
2
 &&
3
 &&
6
 &&
4 &&
 2
 &
 \rule[-1.5ex]{0pt}{4ex} \\
     \hline
 $[1,1;1,1]$
 &&&&
 2
 &&
 4
 &&
6
 &&
4 && 2
   \rule[-1.5ex]{0pt}{4ex}\\
    \hline
  \end{tabular}
      \caption{\small The lowest scalars and their masses.}
\label{scalars0}
\end{table}

For illustration let us consider a scalar field, i.e.~a singlet under the
Lorentz group. The above algorithm gives rise to a Kaluza-Klein tower
\bea
(J,J')~=~(0,0) &\longrightarrow&
\sum_{j,j'}\,[j,j';j,j'] \;,
\eea
built from $SO(4)$ representations which we label by their spins
$[j_{L},j_{L}';j_{R},j_{R}']$ according to \Ref{iso}.
Explicitly, this corresponds to an expansion
\bea
\Phi(z,x,y) &=& \sum_{j,j'}
\phi_{[j,j';j,j']}(z)\,X^{[j,j]}(x)\,Y^{[j',j']}(y)
\;,
\label{ExpansionScalar}
\eea
where the sphere functions $X^{[j,j]}(x)$, $Y^{[j',j']}(y)$ are explicitly given as
symmetric traceless products of~\Ref{XAYA}.

Similarly, a vector say on the first $S^{3}$ gives rise to the Kaluza-Klein towers
\bea
(J,J')=(1,0) &\longrightarrow&
\sum_{j>0,\,j'}\,[j,j';j,j'] +
\sum_{j,j'}\,[j\!+\!1,j';j,j'] +
\sum_{j,j'}\,[j,j';j\!+\!1,j']
\;,\nonumber\\
\label{ExpansionVector}
\eea
and so on. Applying the algorithm to the full spectrum of maximal
nine-dimensional supergravity
leads to the final result~\cite{deBoer:1999rh}
\begin{eqnarray}\label{towerA}
  && \bigoplus_{\ell\ge0,\,\ell'\ge1/2}
  (\ell,\ell';\ell,\ell')_{\rm S}~~\oplus
  \bigoplus_{\ell\ge 1/2,\,\ell'\ge 0}
  (\ell,\ell';\ell,\ell')_{\rm S} \nonumber\\[1ex]
  &&
  \qquad\qquad \oplus\bigoplus_{\ell,\,\ell'\ge 0}
  \big( (\ell,\ell';\ell\!+\ft12,\ell'\!+\ft12)_{\rm S}
  \oplus (\ell\!+\ft12,\ell'\!+\ft12;\ell,\ell')_{\rm S} \big)
  \;,
 \end{eqnarray}
where the fields have already been assembled into supermultiplets
of the supergroup $D^1(2,1;\alpha)_{L} \times D^1(2,1;\alpha)_{R}$
as discussed in chapter~\ref{sec:KKspectrum}.

As an illustration we collect in Table~\ref{statesIIb} for all the
ten-dimensional bosonic degrees of freedom
the lowest $SO(4)\times SO(4)'$ KK states that appear
in their KK decomposition~(\ref{KKdec}). Here
$h$, $(\phi, c_{0})$, $c_{(2)}$, and $c_{(4)}$ denote the fluctuations of the metric,
the scalars, the 2-forms and the 4-form, respectively.
Indices $\mu, \nu, \dots$ label $AdS_{3}$, $m, n,\dots$
and $\bar{m}, \bar{n},\dots$ label the two spheres.
We have omitted all components which do not give rise to
propagating degrees of freedom on $AdS_3$,
in particular half of the self-dual 4-form.

{
\begin{table}[bt]
 \centering
{\footnotesize
  \begin{tabular}{|c||c|c|c|c|c|c|c|c|c|} \hline
    & \!\!$(0,\!0;0,\!0)$\!\! & \!\!$(\ft12,\!0;\ft12,\!0)$\!\! &\!\!$(0,\!\ft12;0,\!\ft12)$\!\!&
  \!\!$(\ft12,\!\ft12;\ft12,\!\ft12)$\!\! & \!\!$(1,\!0;1,\!0)$\!\!
   & \!\!$(0,\!1;0,\!1)$\!\! & \!\!$(\ft12,\!1;\ft12,\!1)$\!\!& \!\!$(1,\!\ft12;1,\!\ft12)$\!\! & \!\!$(1,\!1;1,\!1)$\!\!
   \rule[-2ex]{0pt}{5.5ex}\\
    \hline\hline
   $h_{\mu\nu}$ & &+ &+ &+ & + &+ &+ & + & + \rule[-1.5ex]{0pt}{4ex}\\
   $h_{mn}$ & & & & & + & & & + & + \rule[-1.5ex]{0pt}{4ex}\\
   $h^m_m$ & + & + & + & + & + & + & + & + & +
   \rule[-1.5ex]{0pt}{3.5ex}\\
   $h_{\bar{m}\bar{n}}$ & & & & & & + & + & & + \rule[-1.5ex]{0pt}{3.5ex}\\
   $h^{\bar{m}}_{\bar{m}}$ & + & + & + & + & + & + & + & + & +
   \rule[-1.5ex]{0pt}{4ex}\\
   $h_{\mu m}$ & & + & & + & + & & + & + & + \rule[-1.5ex]{0pt}{3.5ex}\\
   $h_{\mu\bar{m}}$ & & & + & + & & + & + &
   + & +  \rule[-1.5ex]{0pt}{3.5ex}\\
   $h_{m\bar{n}}$ & & & &  + & & & + & + & + \rule[-1.5ex]{0pt}{3.5ex}\\
   $h_{\mu 9}$ & + & + & + & + & + & + & + & + & +
    \rule[-1.5ex]{0pt}{3.5ex}\\
   $h_{m 9}$ & & + & & + & + & & + & + & + \rule[-1.5ex]{0pt}{4ex}\\
   $h_{\bar{m} 9}$ & & & + & + & & + & + & + & + \rule[-1.5ex]{0pt}{3.5ex}\\
   $h_{99}$ & + & + & + & + & + & + & + & + & + \rule[-1.5ex]{0pt}{3.5ex}\\
    \hline
   $\phi, c_0$ & + & + & + & + & + & + & + & + & +
   \rule[-1.5ex]{0pt}{4ex}\\ \hline
   $c_{\mu m}$ & & + & & + & + & & + & + & +
   \rule[-1.5ex]{0pt}{3.5ex}\\
   $c_{\mu\bar{m}}$ & & & + & + & & + & + & + & + \rule[-1.5ex]{0pt}{3.5ex}\\
   $c_{mn}$ & & + & & + & + & & + & + & + \\
   $c_{\bar{m}\bar{n}}$ & & & + & + & & + & + & + & + \\
   $c_{m\bar{n}}$ & & & & + & & & + & + & + \\
   $c_{\mu 9}$ & + & + & + & + & + & + & + & + & + \\
   $c_{m 9}$ & & + & & + & + & & + & + & + \\
   $c_{\bar{m}9}$ & & & + & + & & + & + & + & + \\ \hline
   $c_{\mu mnk}$ & + & + & + & + & + & + & + & + & + \\
   $c_{\mu\bar{m}\bar{n}\bar{k}}$ & + & + & + & + & + & + & + & + & + \\
   $c_{\mu mn\bar{k}}$ & & & & + & & & + & + & + \\
   $c_{\mu m\bar{n}\bar{k}}$ & & & & + & & & + & + & + \\
   $c_{mnk\bar{l}}$ & & & + & + & & + & + & + & + \\
   $c_{mn\bar{k}\bar{l}}$ & & & & + & & & + & + & + \\
   $c_{m\bar{n}\bar{k}\bar{l}}$ & & + & & + & + & & + & + & +  \\
    \hline
  \end{tabular}
  }
 \caption{\small The lowest states in the KK decomposition of the IIB fields.}
\label{statesIIb}
\end{table}
}

Another interesting piece of information is gathered in Table~\ref{scalars0}.
Comparing the field content of Table~\ref{statesIIb} with the supermultiplet
structure from~\Ref{towerA}, we have identified the multiplicites of the
lowest scalar representations together with their AdS masses
(expressed in terms of the boundary conformal dimensions $\Delta=1+\sqrt{1+m^{2}}$).

\end{appendix}

\bibliographystyle{Jopt2}
\bibliography{refss3}

\end{document}